\documentclass[a4paper,11pt]{article}
\usepackage{jheppub} 
\usepackage[T1]{fontenc} 
\usepackage[utf8]{inputenc}
\usepackage{braket}
\usepackage{cancel}
\usepackage{hhline}

\usepackage{mathtools} 
\usepackage{mathrsfs}  

\usepackage{subcaption} 
\usepackage{cleveref} 
\usepackage{slashed} 
\usepackage{physics}
\usepackage{placeins} 
\usepackage[normalem]{ulem}

\def\nn{\nonumber}
\newcommand{\mDm}{m_\chi}

\newcommand{\g}[1]{\gamma_{#1}} 
\newcommand{\Lag}{\mathcal{L}} 
\newcommand{\hc}{\text{h.c.}} 

\newcommand{\mev}{\,\text{MeV}} 
\newcommand{\gev}{\,\text{GeV}} 
\newcommand{\tev}{\,\text{TeV}} 
\newcommand{\arctanh}[1]{\text{arctanh}\left(#1\right)}
\newcommand{\be}{\begin{equation}}
\newcommand{\ee}{\end{equation}}
\newcommand{\CO}{\mathcal{O}}

\newcommand{\orcid}[1]{\href{https://orcid.org/#1}{\includegraphics[width=12pt]{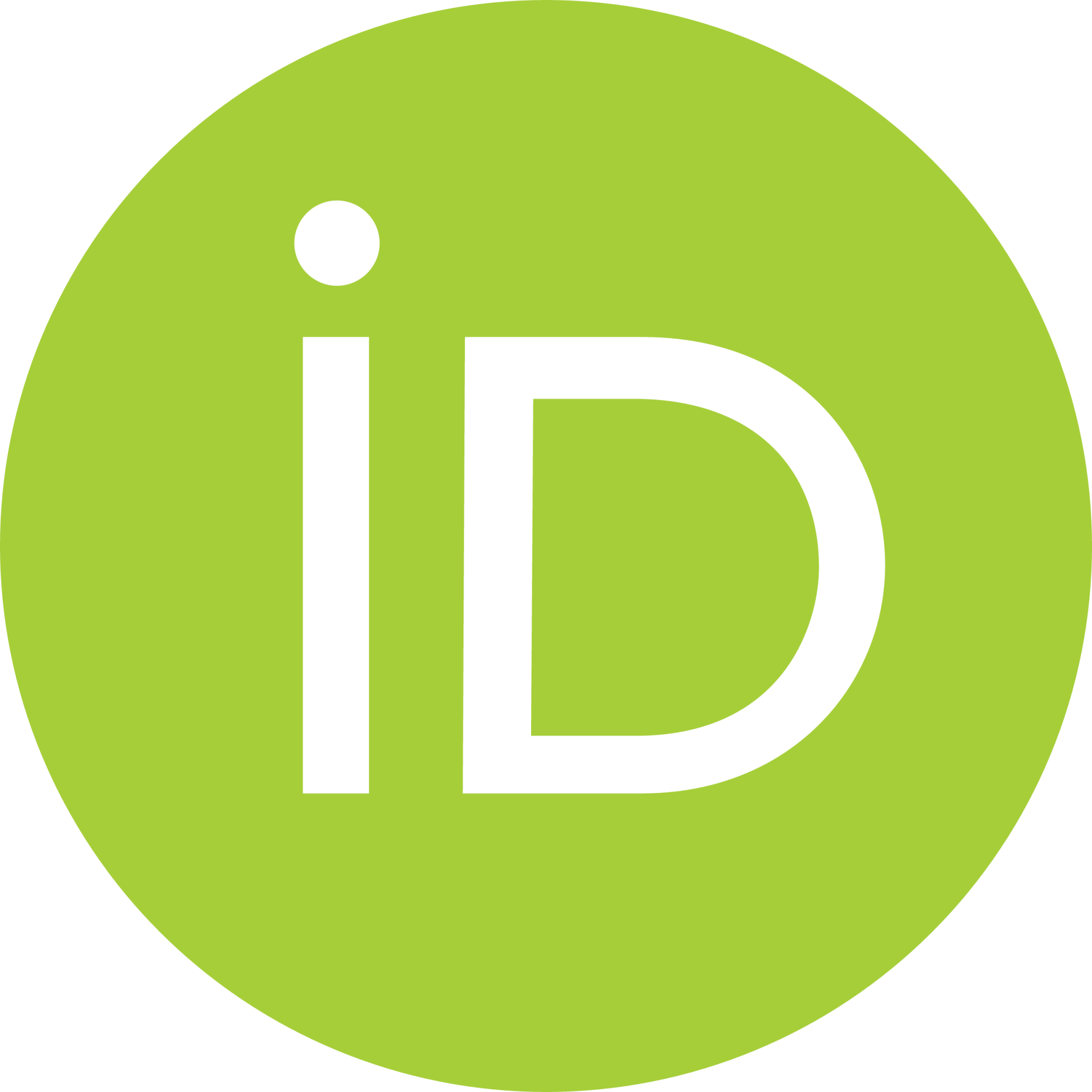}}}
\graphicspath{{./figures/}}
\title{\boldmath Probing Dark Matter Electromagnetic Properties in Direct Detection Experiments}
\author[a]{Alejandro Ibarra\orcid{0000-0001-9935-5247},}
\author[a]{Merlin Reichard\orcid{0000-0002-0568-3272},}
\author[b]{Gaurav Tomar\orcid{0000-0002-3468-5306}}
\affiliation[a]{Technical University of Munich, TUM School of Natural Sciences, Physics Department, 85748 Garching, Germany}
\affiliation[b]{Department of Physics, Indian Institute of Technology Patna, Bihar-801106, India}
\emailAdd{ibarra@tum.de, m.reichard@tum.de, tomar@iitp.ac.in
}
\abstract{Astronomical and cosmological observations indicate that dark matter should interact very weakly with the electromagnetic radiation. Nevertheless, the existence of such interactions is not precluded by observations nor by theoretical considerations. A promising approach to probe the dark matter electromagnetic properties is through the search of photon-mediated dark matter-nucleus interactions in direct detection experiments. In this paper we present a simple methodology to calculate the scattering rate in a direct detection experiment for given values of the dark matter electric charge, charge radius, electric- and magnetic- dipole moments and anapole moment. In our work we include contributions to the scattering from nuclear recoils and from the Migdal effect. We finally apply this formalism to determine exclusion limits on the five electromagnetic interactions using data from XENON1T, LZ, PICO-60 and DS50 experiments, and we discuss the implications for a simplified dark matter model with t-channel mediators.
}
\begin{document} 
\maketitle
\flushbottom

\section{Introduction} 

A vast number of cosmological and astronomical observations suggest that our Galaxy and the Universe at large is filled with a matter component that cannot be accounted for by the Standard Model (SM) of Particle Physics, called dark matter (DM); for reviews, see {\it e.g.} \cite{Jungman:1995df,Bertone:2004pz,Bergstrom:2000pn,Feng:2010gw}. These observations reveal that dark matter does not interact sizeably with the electromagnetic radiation, therefore, in the simplest scenarios the dark matter is regarded as  ``non-luminous''. On the other hand, observations do not preclude the possibility that the dark matter could couple to the electromagnetic radiation. In fact, many well motivated scenarios particle physics scenarios predict that the dark matter field could couple to the photon field through quantum effects~\cite{Pospelov_2000, sigurdson_2004, masso_2009, fitzpatrick_2010, banks_2010, barger_2010, fortin_2011, delNobile_2014, tomar_2022, Bose:2023yll}. More concretely, dark matter particles with spin-1/2 could possess a millicharge  or a charge radius~\cite{berlin_2019, budker_2021}, an electric dipole moment (EDM)~\cite{sigurdson_2004, masso_2009, hambye_2021}, a magnetic dipole moment (MDM)~\cite{sigurdson_2004, masso_2009, delNobile_2014, hisano_2020, hambye_2021}, or the anapole moment~\cite{delNobile_2014, kang_2018} (if the DM is self-conjugate, only the anapole moment remains non-zero). If the dark matter particle has spin-1, additional moments arise~\cite{hisano_2020}.

The dark matter electromagnetic multipoles could lead to observable experimental signatures. In this work we focus on the possible signatures in direct detection experiments, concretely from the coupling of the dark matter particle with the nucleus in the detector material through electromagnetic interactions. The effect of these interactions is two-fold. First, the interaction forces the nucleus to recoil, producing signals in a detector in the form of heat, scintillation and/or ionization. Furthermore, the nuclear recoil could also lead to the ionization of the atom~\cite{bernabei_2007, Ibe:2017yqa,vergados_2005, moustakidis_2005, ejiri_2005, vergados_2013}, through the so-called Migdal effect~\cite{migdal_1941} (efforts to
detect experimentally the Migdal effect have been reported in ~\cite{Xu:2023wev}). The Migdal effect has proven to be particularly valuable for probing dark matter scenarios in the sub-GeV mass range of DM~\cite{LUX:2018akb, EDELWEISS:2019vjv, CDEX:2019hzn, xenon1t_migdal, COSINE-100:2021poy, EDELWEISS:2022ktt, SuperCDMS:2022kgp, ds50_migdal}, where experiments lose sensitivity to the signal from nuclear recoils, and offer a complementary probe to more standard analyses based on electron and nucleus scattering~\cite{Dolan:2017xbu, Essig:2019xkx, Bell:2019egg, GrillidiCortona:2020owp, Flambaum:2020xxo, Knapen:2020aky, Bell:2021zkr, Bell:2021ihi, Wang:2021oha, Liang:2022xbu, Chatterjee:2022gbo, Angevaare:2022zhy, Cox:2022ekg, Li:2022acp, tomar_2022, Kang:2024kec}.

In this paper we extend the formalism presented in \cite{Brenner:2022qku,Brenner:2020mbp} to calculate the scattering rate for a given set of coupling strengths in the non-relativistic effective field theory of dark matter-nucleon interactions, to the case where the scattering is mediated by a photon. We will present simple formulas that relate the millicharge, charge radius, electric and magnetic dipole moment, and anapole moment to the signal rate at a given experiment. For concreteness we will consider the LZ~\cite{LZ:2022lsv}, XENON1T~\cite{XENON:2017vdw}, PICO-60~\cite{PICO:2019vsc}, and DarkSide-50 (DS50)~\cite{DarkSide:2018bpj} experiments. For XENON1T and DS50 (which measure the ionization signal), we will also include the Migdal effect in calculating the signal rate. We will then apply our formalism to derive upper limits on the size of the various dark matter electromagnetic multipoles from the non-observation of a dark-matter induced signal in these experiments. 

To assess the impact of current experiments on concrete models, we will consider a simplified model where the dark matter particle is a spin 1/2 fermion that couples to one Standard Model lepton via scalar mediators. This scenario has been analyzed in recent years, assuming that the dark matter couples only to the left- or the right chirality of the lepton~\cite{Kopp:2014tsa,Garny:2015wea,Ibarra:2015fqa,Sandick:2016zut,Hisano:2018bpz,Baker:2018uox,Kavanagh:2018xeh,Ibarra:2022nzm,Arcadi:2023imv}. In this work, we consider the general case where the dark matter couples to both chiralities, which opens the possibility to CP violating couplings which may induce a dark matter electric dipole moment \cite{Hisano:2018bpz,Herrero-Garcia:2018koq}. 

The paper is organized as follows. In Section
\ref{sec:signals_at_DD_exp} we review the electromagnetic properties of a spin 1/2 fermion and the calculation of the signal rate in a direct detection experiment, due to nuclear recoils and due to the Migdal effect. In Section \ref{subsec:modelin_analysis} we present our formalism to calculate the signal rate for a given set of electromagnetic moments, as well as model independent constraints on these moments, considering one moment at a time as well as when different moments interfere with one another. In Section 
\ref{sec:model} we consider a simplified model of dark matter with a t-channel mediator. We calculate the various electromagnetic moments at the one-loop level and we investigate the prospects to test the model using direct detection experiments, in view of other laboratory constraints. Finally, in Section \ref{sec:conclusions}, we present our conclusions. We also include Appendix \ref{sec:one-loop-calculation} with explicit expressions for the multipole moments in the model presented in Section \ref{sec:model}.

\section{Dark matter detection through photon-mediated interactions with nuclei} 
\label{sec:signals_at_DD_exp}

We consider the scattering of a spin 1/2 dark matter particle $\chi$, with mass $m_\chi$, with a nucleon $N=n,p$, with mass $m_N$, in a target nucleus $T$, with mass $m_T$, in a direct detection experiment mediated by the photon. 
The most general interaction vertex of a spin 1/2 fermion with the electromagnetic field can be cast as~\cite{Nieves:1981zt,Kayser:1982br,Kayser:1983wm,Fukugita:2003en,Nowakowski:2004cv,Broggini:2012df}:
\begin{equation}\label{eq:EMvertex}
	\mathcal{M}^\mu = (\gamma^\mu -q^\mu \slashed{q}/q^2)\left[f_Q(q^2) +f_A(q^2) q^2\g5\right] +i \sigma^{\mu\nu}q_\nu\left[ f_M(q^2) +i f_E(q^2)\g5\right],
\end{equation}
where $f_Q$, $f_A$, $f_M$ and $f_E$ are the charge-, anapole-, magnetic dipole- and electric dipole form factor, and $q$ represents the outgoing four-momentum of the photon. 

The projectiles in the scattering process are dark matter particles in the Milky Way halo, which have a velocity distribution $f(\vec{v}_T)$, with $\vec{v}_T$ the dark matter velocity in the frame of the target. For this, we adopt a standard isotropic Maxwellian in the Galactic rest frame truncated at the escape velocity $u_\text{esc}=550$ km/s, with velocity dispersion 220 km/s, and boosted to the detector frame by the velocity of the Earth. For dark matter particles bound to our Galaxy, the momentum transfer (corresponding to the photon momentum) is $q\lesssim \order{100\mev}$, much smaller than the mass of the target nucleus, so that the charge form factor can be expanded as:
\begin{equation}
    f_Q(q^2) \simeq e Q_\chi + q^2 b_\chi,
\end{equation}
where $Q_\chi$ is the DM electric charge in units of $e$, and $b_\chi$ is the charge radius. On the other hand, for most purposes the remaining form factors can be well approximated by the respective moments, namely the magnetic and electric dipole moments, $\mu_\chi$ and $d_\chi$ respectively, and the anapole moment, $\mathcal{A}_\chi$. 
For a Majorana spin 1/2 fermion, the condition $\chi=\chi^c=i\gamma^2 \chi^*$ implies that all these moments except for the anapole moment must be identically zero.

Keeping operators up to dimension-6, the effective Lagrangian describing the interaction of a spin-1/2 dark matter particle with the photon reads:
\footnote{If the DM electromagnetic moments are generated by new interactions involving light leptons or quarks in the loop, the effective theory could break down and the full momentum dependence ought to be included (see \textit{e.g.} \cite{Kopp:2014tsa}).}
\begin{align}\label{eq:EMLagrangian}
{\cal L}_{\rm int}=
Q_\chi e \bar\chi \gamma^\mu \chi A_\mu +
\frac{\mu_\chi}{2} \bar \chi \sigma^{\mu\nu} \chi F_{\mu\nu} +
\frac{d_\chi}{2} i \bar \chi \sigma^{\mu\nu} \gamma^5 \chi F_{\mu\nu}+
\mathcal{A}_\chi \bar \chi \gamma^\mu \gamma^5 \chi \partial^\nu F_{\mu\nu} +
b_\chi \bar \chi \gamma^\mu \chi \partial^\nu F_{\mu\nu}
\end{align}
The photon, in turn, couples to the nucleons in the nucleus. For small momentum transfer and dark matter-nucleon relative velocity, it was shown in~\cite{Fitzpatrick:2012ix,Anand:2013yka} that the most general dark matter-nucleon interaction Lagrangian can be cast as:
\begin{align}
	\mathcal L = \sum^{15}_{i=1}\sum_{N=n,p} c^{N}_i \mathcal{O}^{N}_i ,
	\label{eq:eff_L}
\end{align}
where $\mathcal{O}_i$ are the non-relativistic effective field theory operators listed in Table~\ref{tab:operators}~\footnote{The operator $\mathcal{O}_2$ is of higher order in $\vec{v}^\perp$, implying a suppressed cross-section for non-relativistic DM. Therefore, we have excluded $\mathcal{O}_2$ from this analysis.} and $c_i^N$ are the coupling strengths to the nucleon $N$. 
For the Lagrangian Eq.~(\ref{eq:EMLagrangian}) the coupling strengths read:
\begin{equation}\label{eq:NR_coefficients}
	\begin{gathered}
			c_1^{N}=  \frac{e^2 Q_N}{q^2} Q_\chi +  \frac{2e Q_N }{4 m_\chi}\mu_\chi +e Q_N  b_\chi\;,\\
			c_4^N= \frac{ 2e g_N }{2 m_N}\mu_\chi  \;,  \quad
			c_5^{N}=  \frac{ 2e Q_N m_N}{q^2} \mu_\chi \;, \quad
			c_6^N= -\frac{2e g_Nm_N}{2 q^2} \mu_\chi  \;,\\
			c_8^{N}=  2e Q_N \mathcal{A}_\chi \;,  \quad
			c_9^{N}= -e g_N \mathcal{A}_\chi  \;, \quad
			c_{11}^N=\frac{2e Q_N m_N }{q^2}  d_\chi \;,
		\end{gathered}    
	\end{equation}
	and all the rest being equal to zero. 
	Here $Q_p = 1, Q_n = 0$ are the proton and neutron electric charge while $g_p = 5.59, g_n = -3.83$ are their $g$-factors.

\begin{table}[t!]
	\begin{center}
		\begin{tabular}{|l|l|}
			\hhline{|-|-|}
			$ \CO_1 = 1_\chi 1_N$ & $\CO_9 = i \vec{S}_\chi \cdot (\vec{S}_N \times \frac{\vec{q}}{m_N})$ \\
			$\CO_3 = i \vec{S}_N \cdot (\frac{\vec{q}}{m_N} \times \vec{v}^\perp)$ & $\CO_{10} = i \vec{S}_N \cdot \frac{\vec{q}}{m_N}$ \\
			$\CO_4 = \vec{S}_\chi \cdot \vec{S}_N$ & $\CO_{11} = i \vec{S}_\chi \cdot \frac{\vec{q}}{m_N}$\\
			$\CO_5 = i \vec{S}_\chi \cdot (\frac{\vec{q}}{m_N} \times \vec{v}^\perp)$ & $\CO_{12} = \vec{S}_\chi \cdot (\vec{S}_N \times \vec{v}^\perp)$ \\
			$\CO_6= (\vec{S}_\chi \cdot \frac{\vec{q}}{m_N}) (\vec{S}_N \cdot \frac{\vec{q}}{m_N})$ & $\CO_{13} =i (\vec{S}_\chi \cdot \vec{v}^\perp  ) (  \vec{S}_N \cdot \frac{\vec{q}}{m_N})$ \\
			$\CO_7 = \vec{S}_N \cdot \vec{v}^\perp$ & $\CO_{14} = i ( \vec{S}_\chi \cdot \frac{\vec{q}}{m_N})(  \vec{S}_N \cdot \vec{v}^\perp )$\\
			$\CO_8 = \vec{S}_\chi \cdot \vec{v}^\perp$ & $\CO_{15} = - ( \vec{S}_\chi \cdot \frac{\vec{q}}{m_N}) \big((\vec{S}_N \times \vec{v}^\perp) \cdot \frac{\vec{q}}{m_N}\big)$ 
			\\ \hline
		\end{tabular}
		\caption{Non-relativistic Galilean invariant operators for DM with spin $1/2$. Here, $\vec{S}_\chi$ and $\vec{S}_N$ represent the spins of the DM and nucleon, respectively, while $\vec{v}^\perp=\vec{v}+\vec{q}/2\mu_{\chi N}$ (where $\mu_{\chi N}$ is the DM-nucleon reduced mass).}
		\label{tab:operators}
	\end{center}
\end{table}

The differential rate of scattering of a dark matter particle with relative velocity $v_T$ producing a nuclear recoil with energy $E_R$ can be calculated from 
\be
\frac{d R_{\chi T}}{d E_R d v_T}=\sum_T N_T\frac{\rho_{\chi}}{m_{\chi}} f(\vec{v}_T) v_T \frac{d\sigma_T}{d E_R},
\label{eq:dr_der_ER_vT}
\ee
where $\rho_{\chi}\simeq 0.3\,{\rm GeV}/{\rm cm}^3$ is the DM mass density in the neighborhood of the Sun, $N_T$ represents the number of the nuclear targets of species $T$ in the detector, and the differential cross-section is:
\be
\frac{d\sigma_T}{d E_R}=\frac{2 m_T}{4\pi v_T^2}\left [ \frac{1}{2 j_{\chi}+1} \frac{1}{2 j_{T}+1}|\mathcal{M}_T|^2 \right ],
\label{eq:dsigma_de}
\ee
where $|\mathcal{M}_T|^2$  is the squared amplitude,
given by \cite{Anand:2013yka}
\begin{equation}
    \frac{1}{2 j_{\chi}+1} \frac{1}{2 j_{T}+1}|\mathcal{M}_T|^2=\frac{4\pi}{2j_T+1}\sum_{\tau=0}^1\sum_{\tau'=0}^1\sum_k R_k^{\tau\tau'}\left[c_i^\tau, c_j^{\tau'}, (v^\perp)^2,\frac{q^2}{m_N^2}\right]W_{Tk}^{\tau\tau'}(y).
\end{equation}
Here, $R_k^{\tau\tau'}$ ($W_{Tk}^{\tau\tau'}$) are the DM (nuclear) response functions, and $y\equiv(qb/2)^2$ with $b$ the size of the nucleus.

Finally, the total scattering rate producing a nuclear recoil with energy $E_R$ is obtained from
\be
\frac{d R_{\chi T}}{d E_R}=\int_{v_\text{min}}d^3 v_T \,
\frac{d R_{\chi T}}{d E_R dv_T},
\label{eq:dr_der}
\ee
with $v_\text{min}=\sqrt{\frac{m_T E_R}{2\mu_T^2}}$ and $\mu_T$ the DM-target nucleus reduced mass. 

The dark matter scattering with a nucleus could also cause the ejection of the electron from the atom, the so-called Migdal effect~\cite{migdal_1941}, producing an electromagnetic signal that could also be detected. The ionization rate generating an energy at the detector $E_{\rm det}$ due to the Migdal effect is calculated as~\cite{Ibe:2017yqa}:
\begin{equation}
	\label{diff_rate_migdal}
	\frac{dR_{\chi T}}{dE_{\rm det}}=\int_0^{\infty} d E_R \int_{v_\text{min}(E_R)}^\infty dv_T \,\frac{d^2R_{\chi T}}{dE_R dv_T}\times \frac{1}{2\pi}\sum_{n,l}\frac{d}{dE_e}p_{nl\rightarrow(E_e)}(q_e).
\end{equation}
Here, $p_{nl\rightarrow(E_e)}(q_e)$ is the probability of ionizing an electron in an orbital with quantum number $n$ and $l$ and de-excitation energy $E_{nl}$,  producing a free electron with energy $E_e$, for an average momentum transfer $q_e$ to an individual electron (given in the rest frame of the target nucleus by $q_e=m_e\sqrt{2E_R/m_T}$). Besides, the differential rate of scattering with the nucleus is given by Eq.~(\ref{eq:dr_der_ER_vT}) and 
\begin{equation}
	v_{\rm min}(E_R)=\frac{m_T E_R+\mu_T E_{\rm EM}}{\mu_T\sqrt{2m_T E_R}}.
	\label{eq:vmin}
\end{equation}
Here, $E_{\rm EM} = E_e + E_{nl}$ is the ionization energy deposited in the detector  and $E_R$ is related to the total energy deposited in the detector through $ E_{\rm det} = E_{\rm EM} + Q E_R$, with $Q$ the quenching factor. 

In our analysis, we considered the LZ~\cite{LZ:2022lsv}, XENON1T~\cite{XENON:2017vdw}, PICO-60~\cite{PICO:2019vsc}, and DS50~\cite{DarkSide:2018bpj} direct detection experiments. We utilized the WimPyDD code~\cite{Jeong:2021bpl}, which includes the implementation of XENON1T and PICO-60 experiments in its published version.  Additionally, we utilize the ionization probabilities calculated in Ref.~\cite{Ibe:2017yqa}.~\footnote{Recently, ionization probabilities have been calculated using the Dirac-Hartree-Fock approach~\cite{Cox:2022ekg}, differing from the dipole approximation pursued in~\cite{Ibe:2017yqa}. However, in the energy ranges relevant to our investigation, both methods yield similar results.}
The LZ and DS50 experiments have been implemented in WimPyDD to perform our analysis. We used the isolated atom approximation, where we focus solely on the contributions from inner shell electrons while disregarding the valence electrons. This approach is commonly adopted as the inner shell electrons are the primary contributors. However, in the case of liquid detectors such as XENON1T~\cite{xenon1t_migdal} and DS50~\cite{ds50_migdal}, the presence of atoms in a liquid introduces a shift in electronic energy levels. Nevertheless, this effect is of lesser significance for inner shell electrons. 

\begin{figure}[t!]
    \centering
		\includegraphics[width=0.49\textwidth]{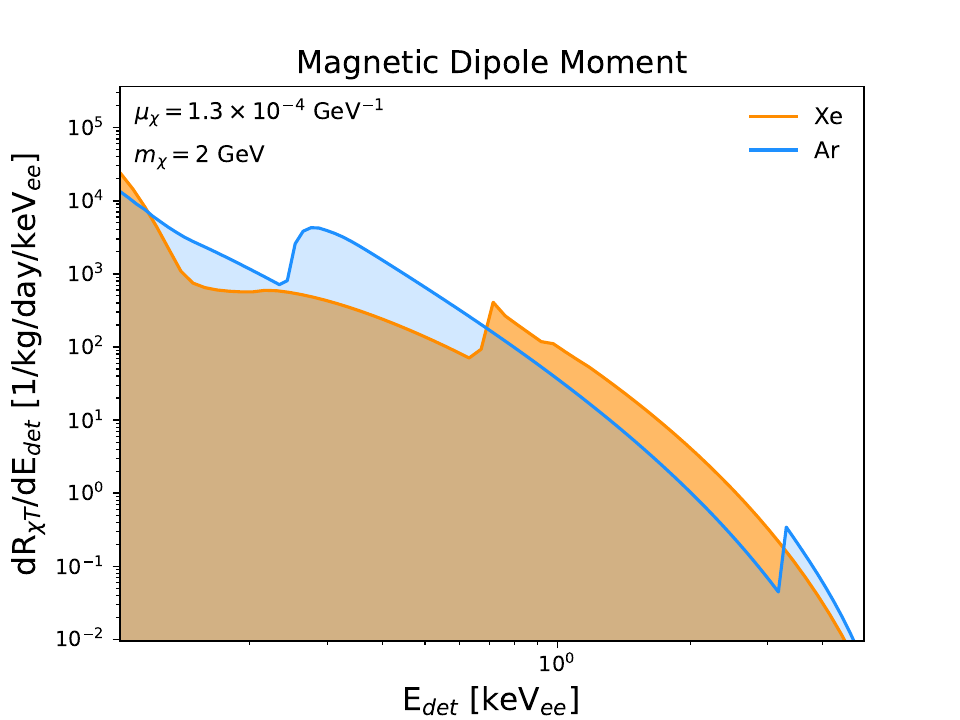}
		\includegraphics[width=0.49\textwidth]{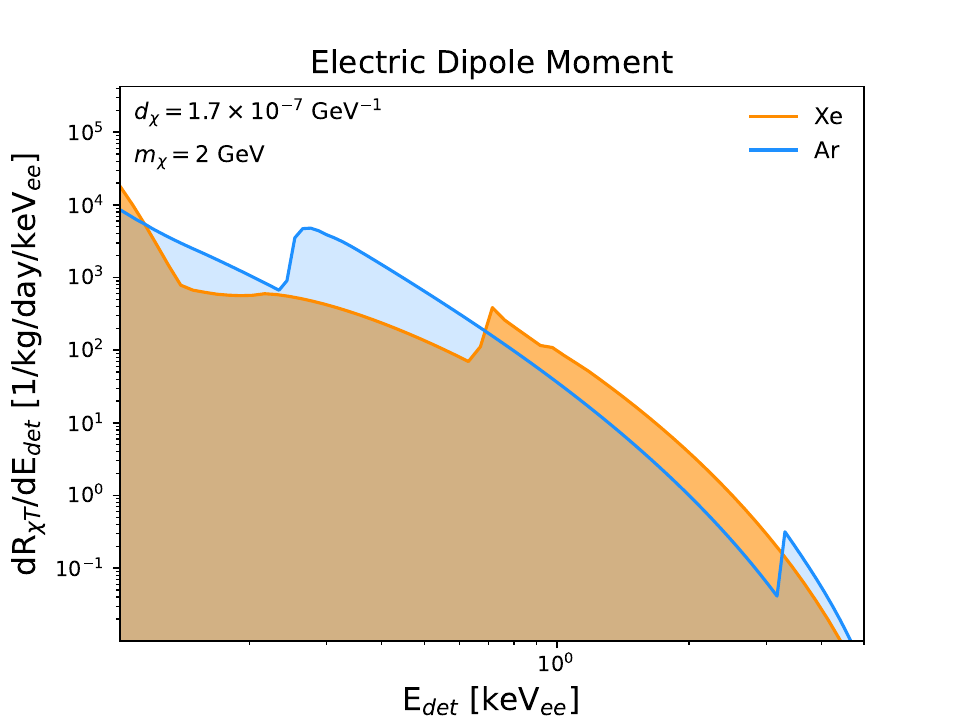}\\
		\includegraphics[width=0.49\textwidth]{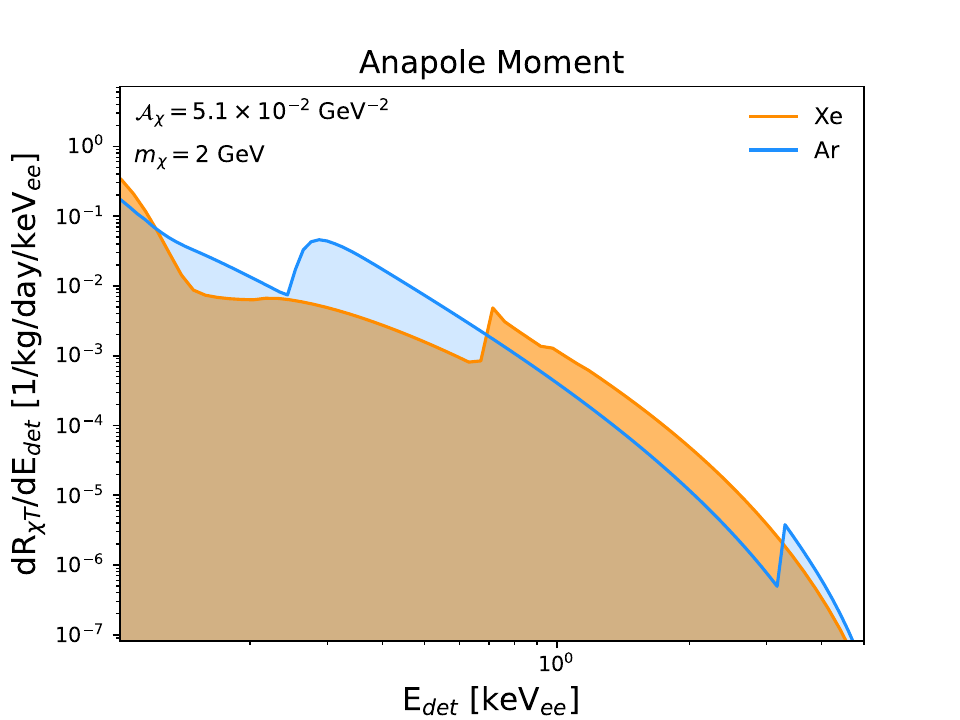}
		\includegraphics[width=0.49\textwidth]{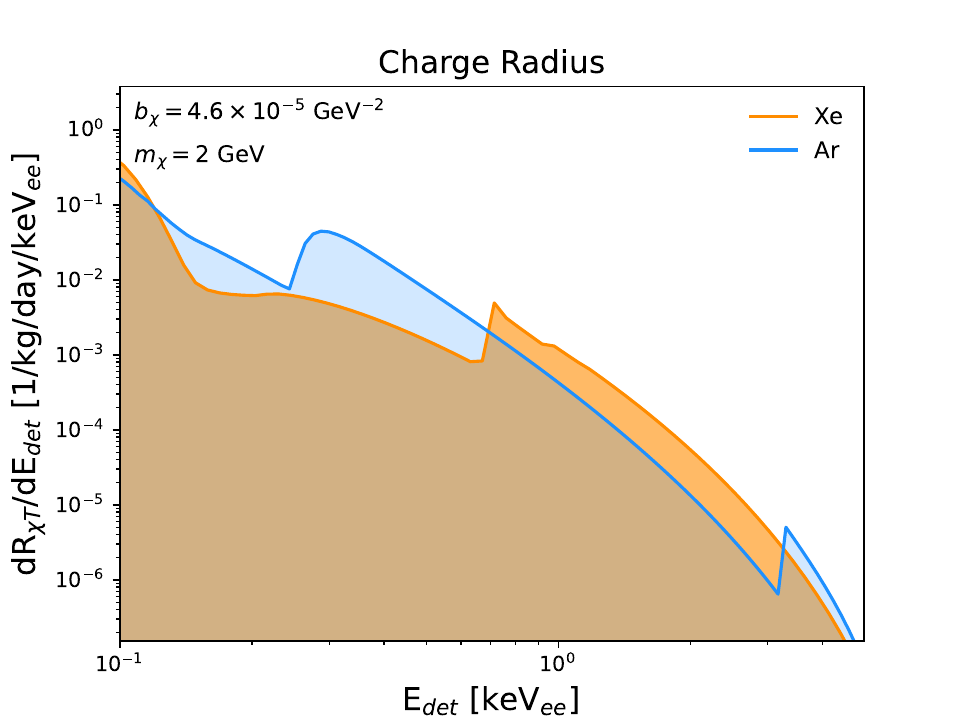}
		\includegraphics[width=0.49\textwidth]{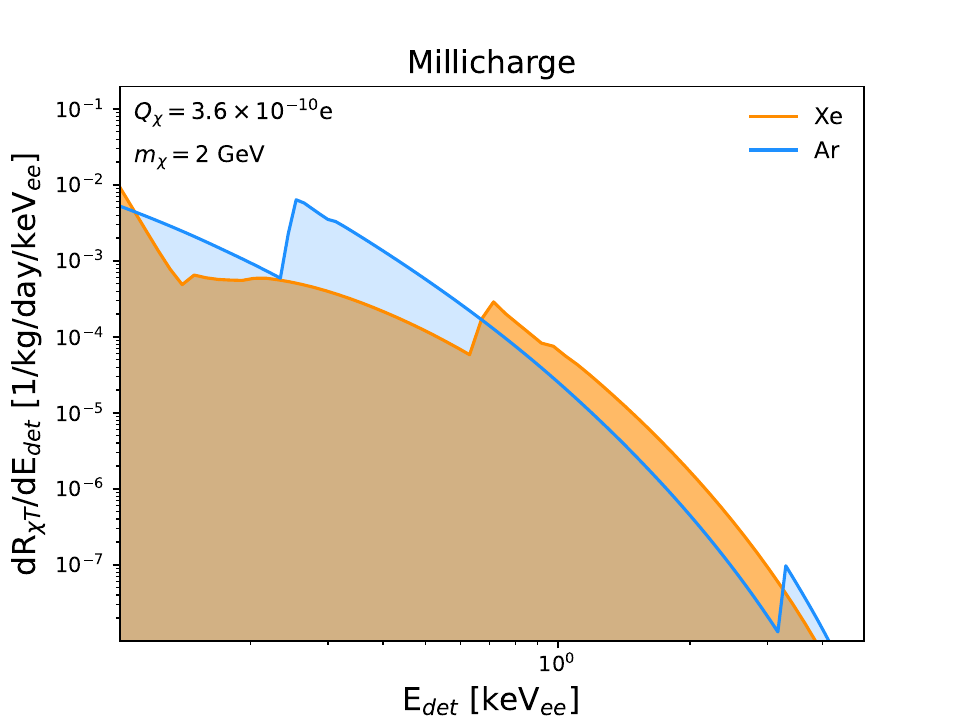}
	\caption{
		Migdal differential rate as a function of the detected energy for $m_\chi=2$ GeV, for a xenon and argon target, assuming that the DM interacts with the nucleus through the magnetic dipole, electric dipole, anapole, charge radius, or millicharge interaction.}
	\label{fig:diff_rate}
\end{figure}

We show in  Fig.~\ref{fig:diff_rate} the differential event rate for the Migdal effect for $m_\chi=2$ GeV in a xenon and argon target, as a function of the detected energy, $ E_{\rm det}$, for representative values of the dark matter magnetic dipole, electric dipole, anapole, charge radius, and millicharge.  The different peaks correspond to the differential rates
of the shells n = 1, 2, 3, 4, and 5, depending on the considered targets, which are summed over.
The total signal rate induced by the Migdal effect is typically negligible compared to the more studied signals from elastic nuclear recoils, unless for very low DM masses, for which the latter is below the experimental threshold.

\section{Model-independent approach}
\label{subsec:modelin_analysis}
Since the event rate is a quadratic form of the coupling strengths of the effective theory, it is convenient to cast the total number of signal events in a given experiment  $\mathcal{E}$ as~\cite{Brenner:2022qku,Brenner:2020mbp},
\begin{align}\label{eq:signal_rate_master_eq}
\mathcal{N}_{\rm sig}^{\cal E}= \begin{pmatrix} Q_\chi & \mu_\chi & d_\chi & {\cal A}_\chi & b_\chi \end{pmatrix}
\begin{pmatrix} 
\mathbb{N}^{\cal E}_{Q_\chi Q_\chi} & \mathbb{N}^{\cal E}_{Q_\chi \mu_\chi}& 0 & 0 & \mathbb{N}^{\cal E}_{Q_\chi b_\chi} \\
\mathbb{N}^{\cal E}_{Q_\chi \mu_\chi} & \mathbb{N}^{\cal E}_{\mu_\chi \mu_\chi} & 0 & 0 & \mathbb{N}^{\cal E}_{\mu_\chi b_\chi} \\
0 & 0 & \mathbb{N}^{\cal E}_{d_\chi d_\chi} & 0 & 0 \\
0 & 0 & 0 & \mathbb{N}^{\cal E}_{\mathcal{A}_{\chi} \mathcal{A}_\chi} & 0 \\
\mathbb{N}^{\cal E}_{Q_\chi b_\chi} & \mathbb{N}^{\cal E}_{\mu_\chi b_\chi} & 0 & 0 & \mathbb{N}^{\cal E}_{b_\chi b_\chi} \end{pmatrix}
\begin{pmatrix} Q_\chi \\ \mu_\chi \\ d_\chi \\ {\cal A}_\chi \\ b_\chi \end{pmatrix},
\end{align}
where the matrix elements $\mathbb{N}_{ij}^\mathcal{E}$ are functions of the dark matter mass, and encode all detector specifics, and astrophysical inputs ({\it i.e.} local DM density  and velocity distribution). Note the possible interference among the  millicharge, magnetic dipole, and charge radius interactions, which is due to the simultaneous contribution of all these moments to the coupling strength $c_1^N$, {\it cf.} Eq.~(\ref{eq:NR_coefficients}).

\begin{figure}[t!]
    \centering
        \includegraphics[width=0.49\linewidth]{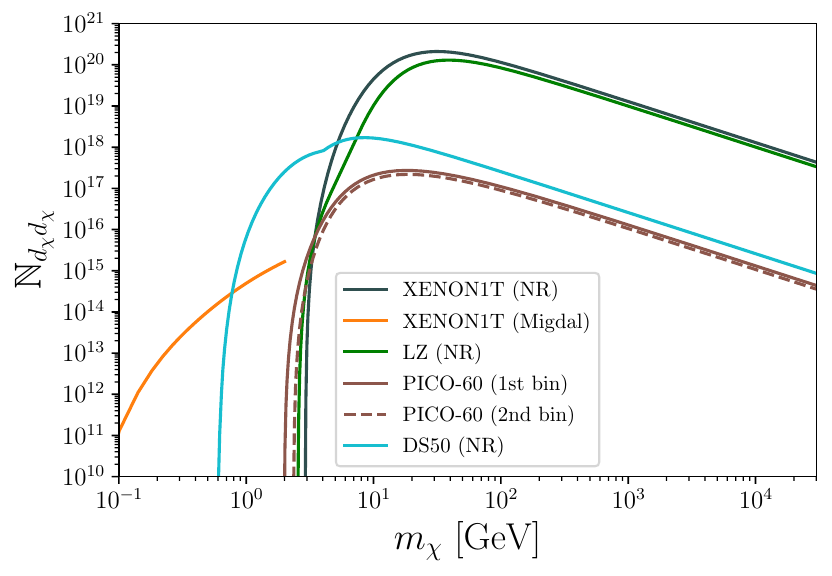}
        \includegraphics[width=0.49\linewidth]{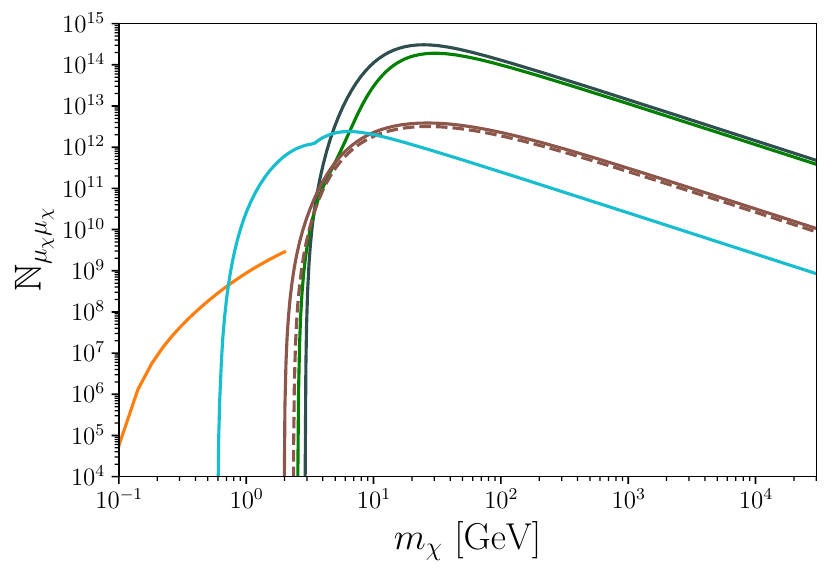}\\
        \includegraphics[width=0.49\linewidth]{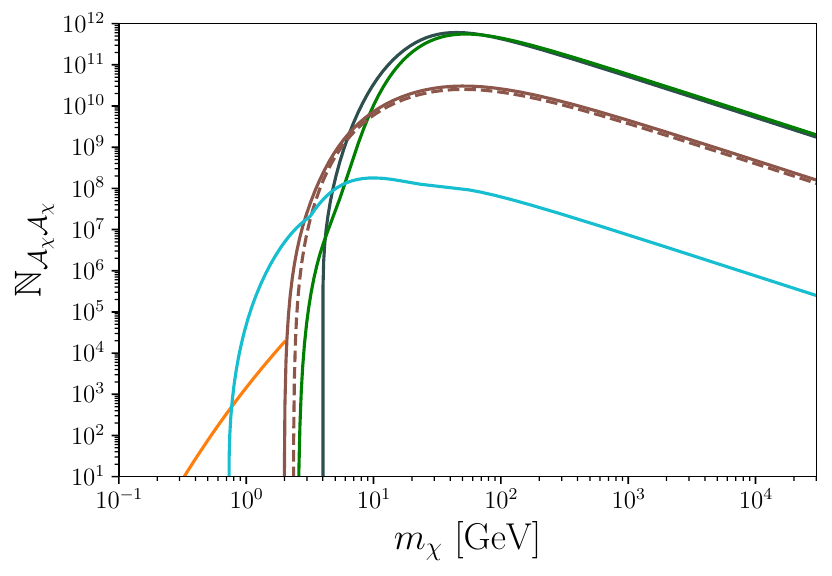}
        \includegraphics[width=0.49\linewidth]{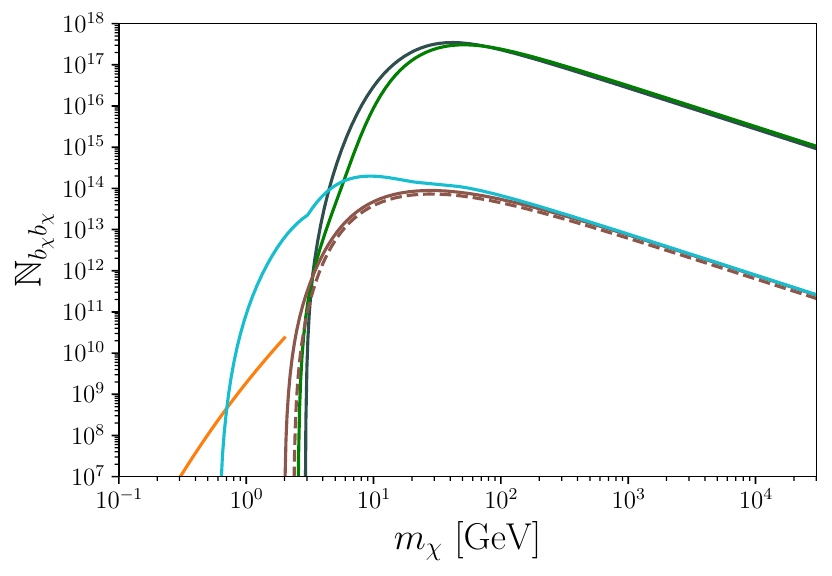} \\
        \includegraphics[width=0.49\linewidth]{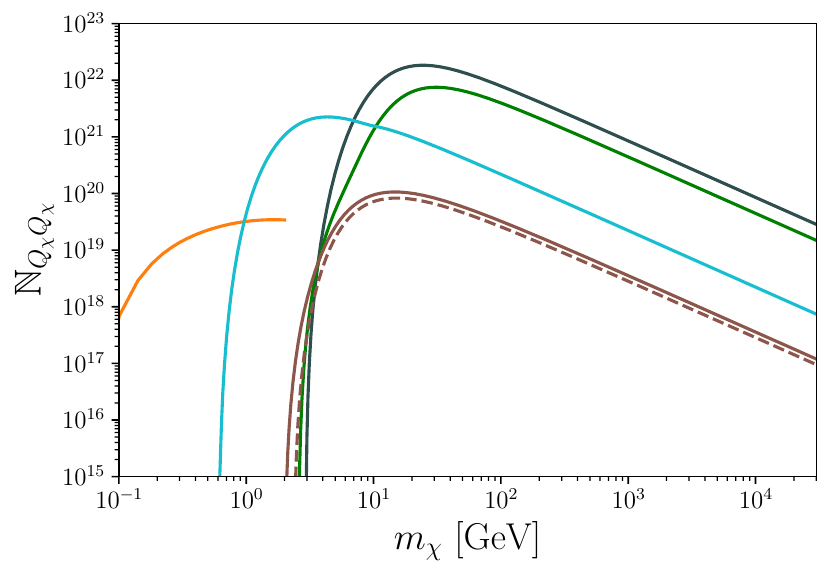} 
        \includegraphics[width=0.49\linewidth]{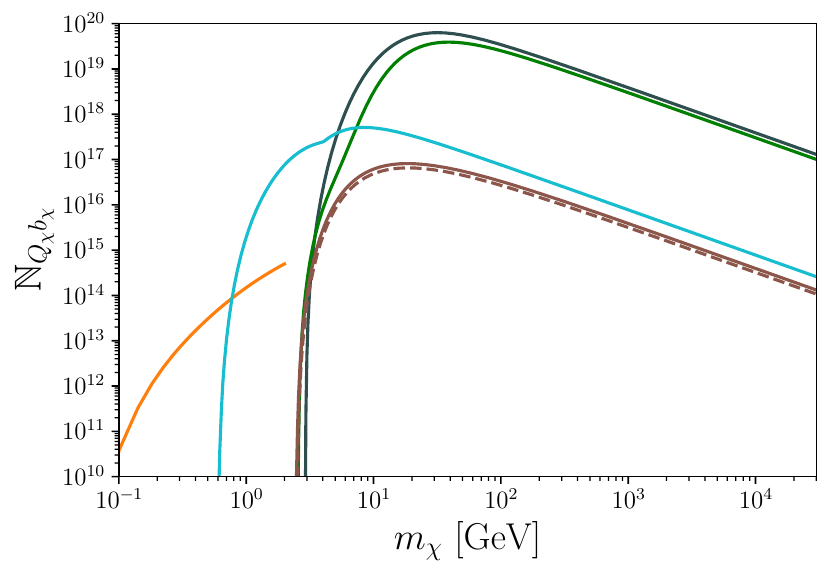}\\
        \includegraphics[width=0.49\linewidth]{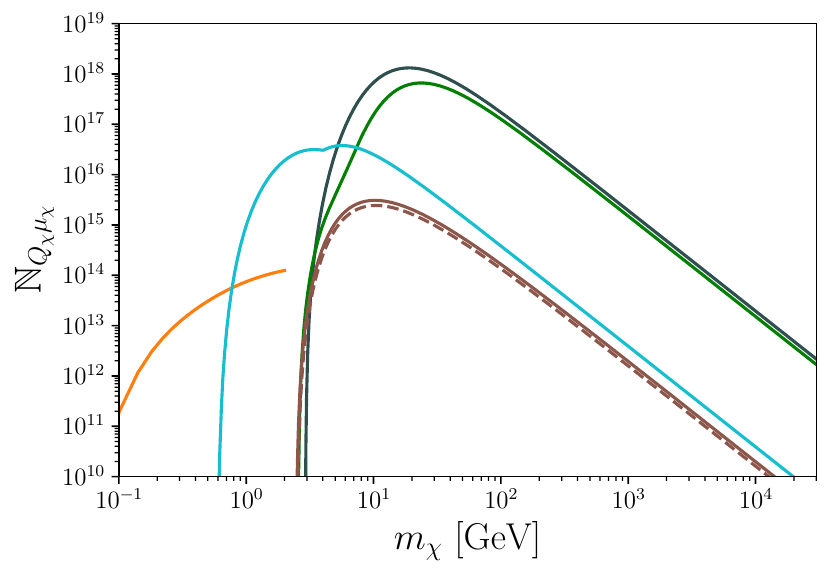}
        \includegraphics[width=0.49\linewidth]{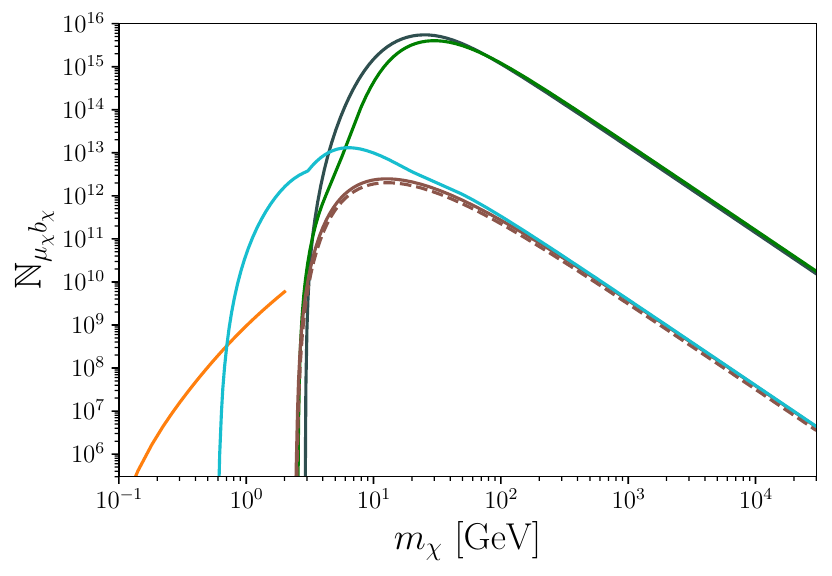}
    \caption{Elements of the matrix $\mathbb{N}_{ij}^\mathcal{E}$ defined in Eq.~(\ref{eq:signal_rate_master_eq}) for different experiments $\mathcal{E}$.}
    \label{fig:Rate_functions_1}
\end{figure}

We have calculated the  matrix elements $\mathbb{N}_{ij}^\mathcal{E}$ using Eqs.~(\ref{eq:dr_der}) and~(\ref{diff_rate_migdal}) for DM-nucleus elastic scattering and for Migdal scattering, respectively, for the experiments LZ, XENON1T, PICO-60, and DS50, and are shown for reference in Figs. \ref{fig:Rate_functions_1}.\footnote{The $\mathbb{N}^\mathcal{E}$-matrices can be downloaded from \href{https://github.com/ga42puq/EMmoments-rate-matrices}{https://github.com/ga42puq/EMmoments-rate-matrices}.} For DS50 only the largest value among all bins considered is shown for simplicity. 

This formalism can be straightforwardly applied to derive model-independent upper limits on the various electromagnetic moments from the non-observation of a signal in these experiments. For the LZ experiment, we used an exposure of $3.3\times 10^5$ kg-days while assuming no signal in the lower half of the nuclear recoil band below the red curve of Fig. 1 of Ref.~\cite{LZ:2022lsv} in the range of 2 PE $\leq S_1\leq 70$ PE. Regarding the DS50 experiment~\cite{DarkSide:2018bpj}, we subtracted the background by minimizing a likelihood function. Further details about its implementation are discussed in Appendix B of~\cite{Kang:2018odb}.

In our Migdal analysis of XENON1T~\cite{xenon1t_migdal}, we assume 49 DM events from an exposure of 22 tonne-days in the range $0.186~{\rm keVee}\leq E_{det} \leq 3.8~{\rm keVee}$, corresponding at a $90\%$ confidence level to the 61 observed events. As pointed out~\cite{tomar_2022}, reproducing the profile–likelihood analysis used by DS50~\cite{ds50_migdal} is difficult, but it is noticed that the Migdal energy spectrum $dR_{\chi T}/dE_{det}$ is fixed by the ionization probabilities $p({q_e})$ and is the same for all interactions. Therefore, we used the normalization of the exclusion plot in \cite{ds50_migdal} to estimate the upper bound on the DM events for all interactions. We used an exposure of 12.5 tonne-days and energy bins of $0.083 \leq E_{det} \leq 0.106$ keVee, with 20 events, and $1.4~{\rm keVee} \leq E_{det} \leq 10~{\rm keVee}$ with no events, to reproduce the exclusion limit of Fig. 3 in~\cite{ds50_migdal}.

\begin{figure}[t!]
    \centering
        \includegraphics[width=0.49\linewidth]{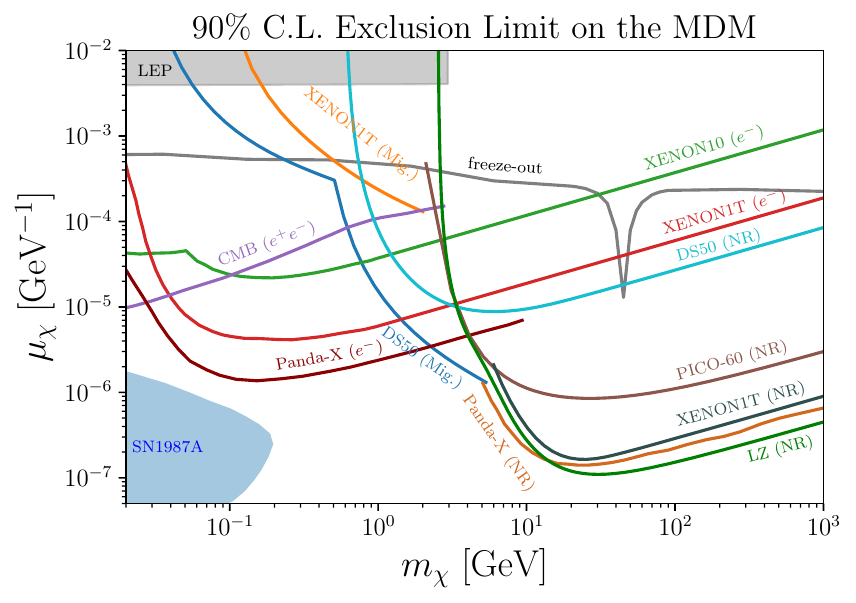}
        \includegraphics[width=0.49\linewidth]{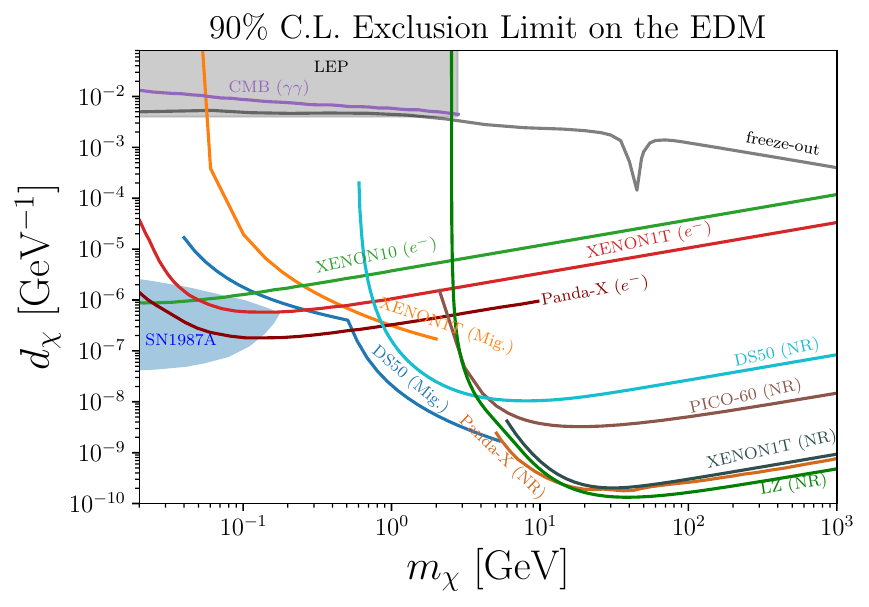}\\
        \includegraphics[width=0.49\linewidth]{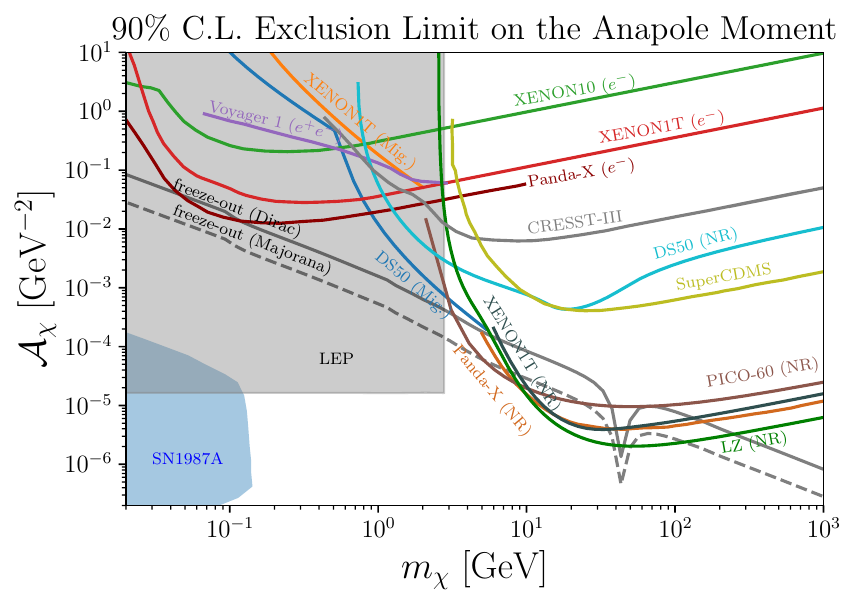}
        \includegraphics[width=0.49\linewidth]{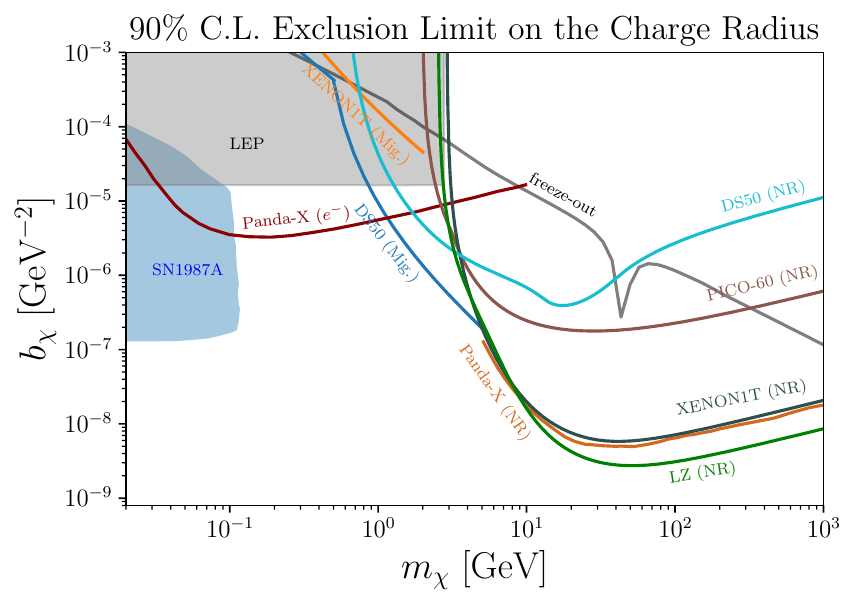} \\
        \includegraphics[width=0.49\linewidth]{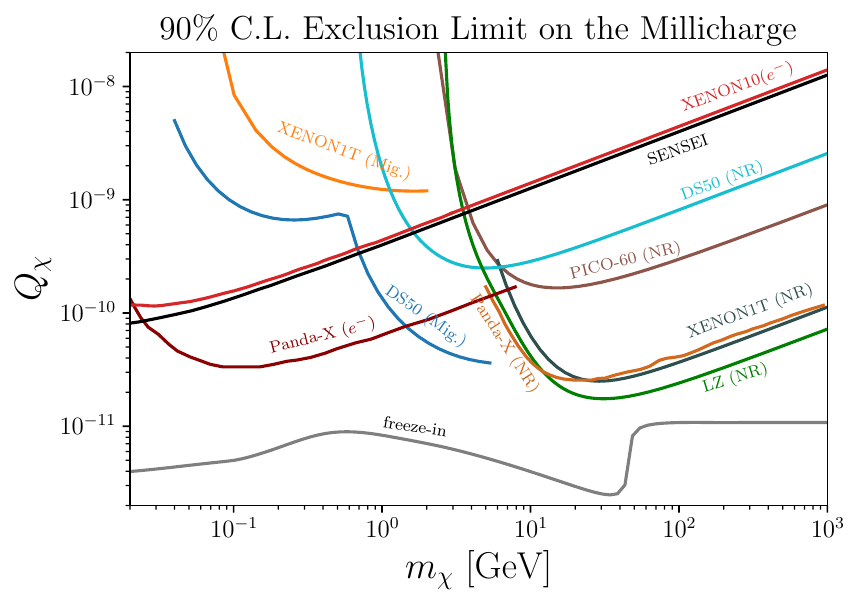}
    \caption{$90\%$ C.L. limits on the electromagnetic interactions of spin-1/2 dark matter. We also show the values of the corresponding moment leading to the correct dark matter abundance via freeze-out/freeze-in. See main text for details.}
    \label{fig:ExclusionLimits}
\end{figure}

The upper limits on various electromagnetic moments, assuming only one interaction at a time, are shown  in Fig.~\ref{fig:ExclusionLimits}. 
The plot also shows the constraints on these moments from LEP, Supernova 1987A, CMB and Voyager \cite{Chu:2018qrm}, as well as from the non-detection of DM-induced electron recoils at XENON10/XENON1T \cite{Catena:2019gfa} and PANDA-X \cite{Liang:2024ecw}.
The constraints from LEP are not fully model-independent since it is implicitly assumed that the scales generating the moments are larger than the collider center-of-mass energies. Additional nuclear recoil constraints from PANDA-X \cite{PandaX:2023toi}, and for the anapole operator from SuperCDMS and CRESST-III \cite{Ibarra:2022nzm}, are shown.
For the millicharge we show limits from nuclear recoil direct detection experiments \cite{Belanger:2020gnr}. We recast the XENON10 electron recoil limits from \cite{Essig:2017kqs} and from SENSEI \cite{SENSEI:2020dpa} assuming a massless dark photon generating the millicharge \cite{Emken:2019tni}.

We also show the values needed for a thermal DM candidate to reach the measured relic abundance of $\Omega h^2\simeq 0.12$ \cite{Planck:2018vyg} within the standard freeze-out (freeze-in for the millicharge) paradigm, obtained via \texttt{FeynRules} \cite{Christensen:2008py} and \texttt{micrOMEGAs} \cite{Belanger:2013oya,Belanger:2018ccd}. Note that we use a model in which DM couples to the hypercharge gauge boson $B^\mu$ instead of the SM photon field $A^\mu$ only, as the latter leads to a non-unitary annihilation cross section into pairs of W bosons \cite{Arina:2020tuw,Choi:2024uva}. In this scenario, also Z-boson mediated processes enter the DM annihilation cross section, which are enhanced for $\mDm \simeq m_Z/2$ and result in the dip in the freeze-out lines observed in Fig.~\ref{fig:ExclusionLimits}. Similarly, for the millicharge scenario, the additional Z-mediated process leads to an increase in the annihilation cross section $\sigma(\text{SM}\,\text{SM}\rightarrow \chi\,\chi)$ if kinematically allowed. In contrast to the photon-mediated approach, this leads to the valley-shape of the coupling $Q_\chi$ needed to reach $\Omega h^2 \simeq 0.12$ in the region $1\gev \lesssim m_\chi \lesssim m_Z/2$.

We find that the nuclear recoil limits dominate the exclusion plots for $\mDm \gtrsim 10\gev$ for every effective coupling considered in this work. Only for the electric dipole, the XENON1T (Migdal) analysis leads to strong limits around $\mDm \sim 1\gev$, whereas for the other operators, this constraint is sub-dominant. Meanwhile, in the same region, the DS50 (Migdal) limits give the best constraints for all electromagnetic operators, complementing nuclear- and electron recoil limits.

\begin{figure}
    \centering
        \includegraphics[width=0.49\linewidth]{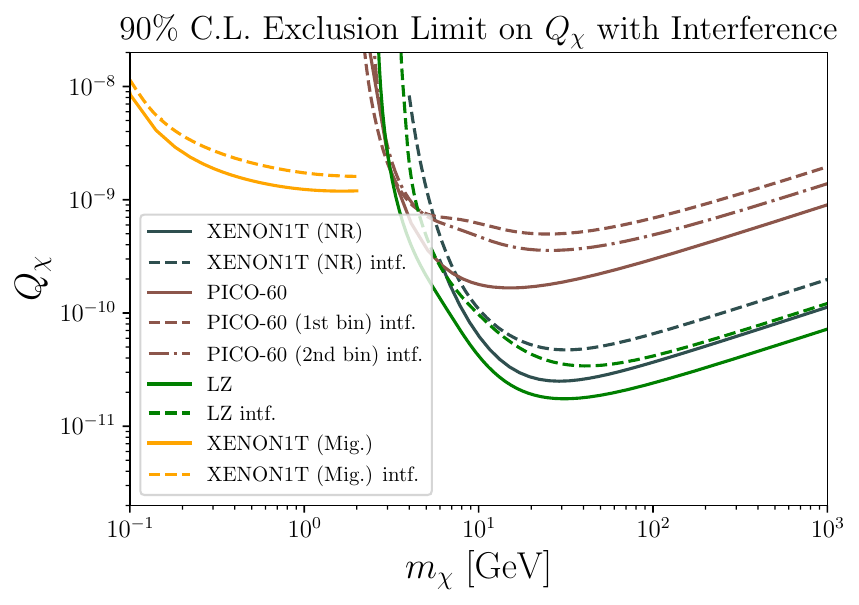}
        \includegraphics[width=0.49\linewidth]{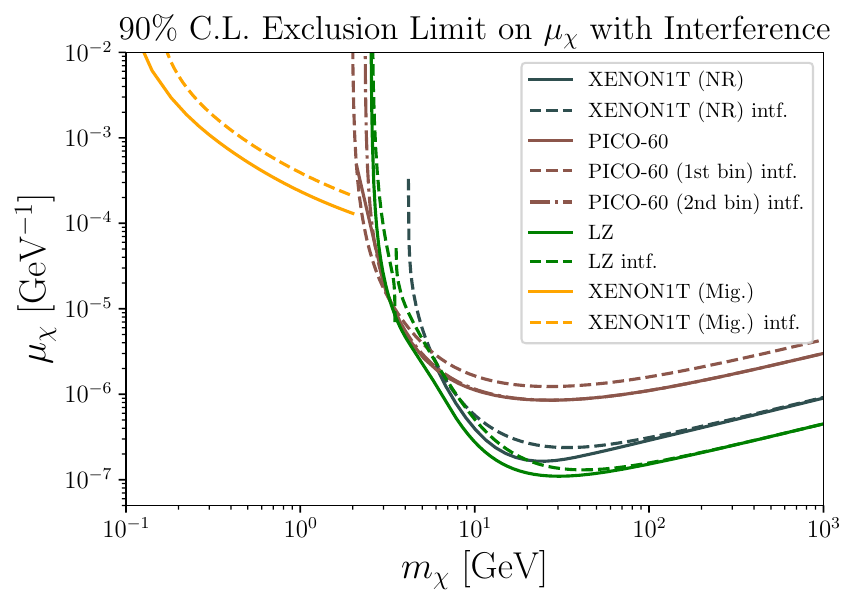} \\
        \includegraphics[width=0.49\linewidth]{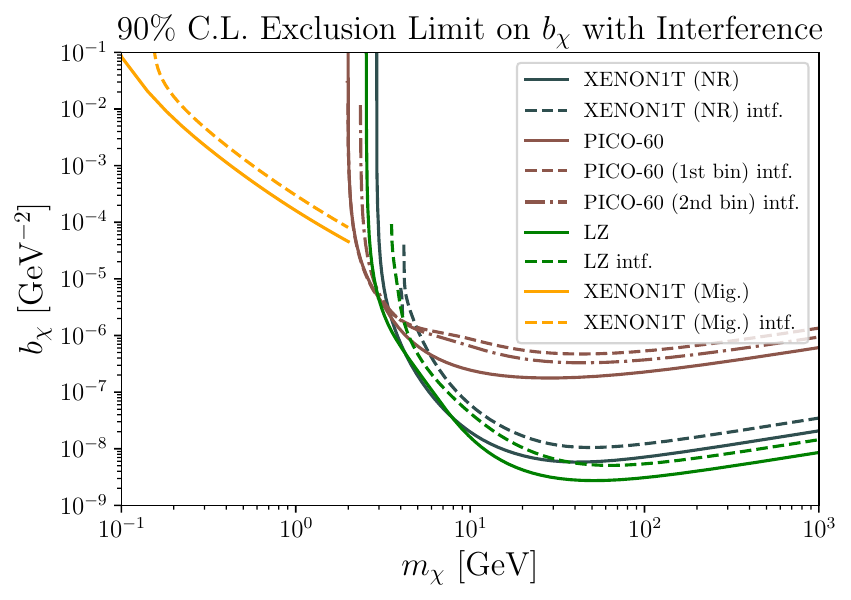}
    \caption{90\% C.L. exclusion limits on the interfering electromagnetic interactions of a Dirac fermionic DM candidate. The continuous lines assume one interaction at a time, while the discontinuous lines include possible interference effects. }
    \label{fig:ConservativeLimitsIncludingInterference}
\end{figure}

In Fig.~\ref{fig:ConservativeLimitsIncludingInterference}, we study the impact of including the interference terms of Eq.~(\ref{eq:signal_rate_master_eq}), following the approach of~\cite{Brenner:2022qku,Brenner:2020mbp}. The continuous lines represent the exclusion limits obtained when considering only the diagonal terms, while the discontinuous lines are the most conservative limits on the moments, and were derived including a possible interference among interactions. For the PICO-60 experiment, we show separately the result from the first- and second bin (dashed and dash-dotted, respectively).

\begin{figure}[t!]
    \centering
        \includegraphics[width=0.49\linewidth]{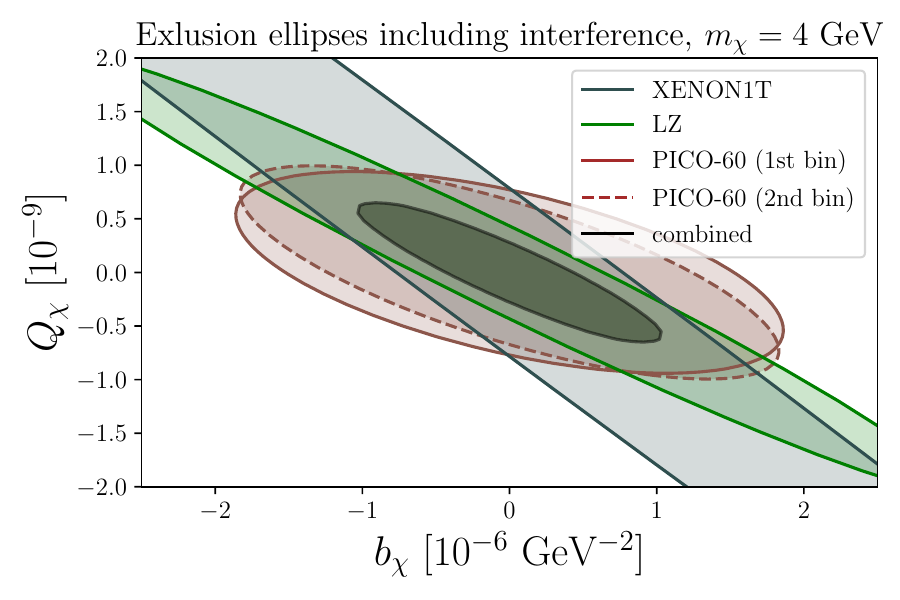}
        \includegraphics[width=0.49\linewidth]{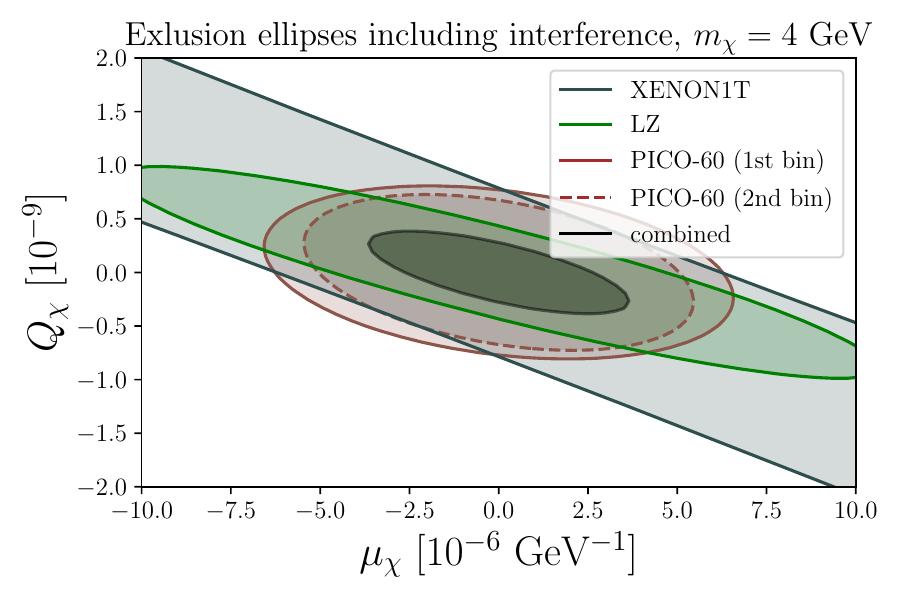} \\
        \includegraphics[width=0.49\linewidth]{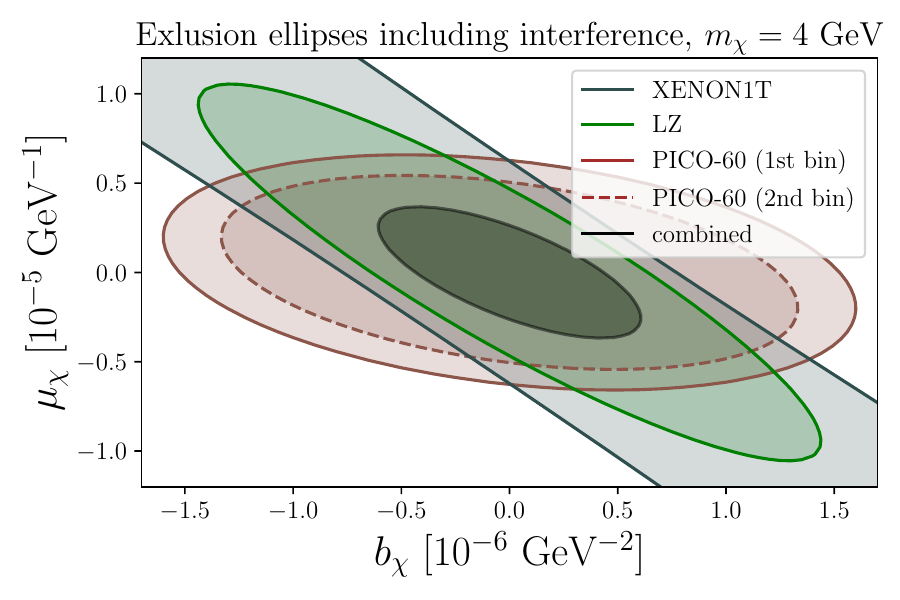}
    \caption{$90\%$ C.L. exclusion regions in the parameter space spanned by two electromagnetic multipoles, including the interference effects, from XENON1T, LZ and PICO-60. We show also in black the exclusion region from combining the three experiments.}
    \label{fig:ExclusionEllpises}
\end{figure}

As the sensitivity to the DM electromagnetic interactions depends on the target material, combining experiments with different targets can improve our understanding of the DM electromagnetic properties. In Fig.~\ref{fig:ExclusionEllpises}, we consider two electromagnetic multipoles at a time, and we plot the exclusion ellipses from the non-observation of nuclear recoils at XENON1T, PICO-60, and LZ for $\mDm = 4\gev$. The tilt of the ellipses in Fig.~\ref{fig:ExclusionEllpises} is due to the presence of interference terms and to the different sensitivity of the experiments to the various electromagnetic multipoles. As apparent from the plot, combining the null-results of different experiments can lead to a significant reduction of the allowed parameter of the model compared to the region excluded by each single experiment. Combining different targets could also be pivotal in interpreting a putative signal (for details, see \cite{Brenner:2022qku}).

\section{Electromagnetic multipoles in t-channel mediator models}
\label{sec:model}

We now particularize the previous model-independent results to a simplified model where the DM particle, $\chi$, is a spin 1/2 fermion. We assume that the DM particle interacts with the standard model (SM) through Yukawa interactions. Concretely, we assume that the DM interacts with a $SU(2)_L$ doublet $F$ (singlet $f_R$) through a complex scalar doublet $S_L$ (singlet $S_R$), with hypercharge opposite to $F$ ($f_R$). Further, we assume the existence of an exact discrete $Z_2$ symmetry under which the dark sector particles $\chi$, $S_L$ and $S_R$ are odd, while the Standard Model particles are even, and which protects the DM against decay. The interaction Lagrangian reads:
\begin{align}\label{eq:toymodel_gauge_interaction}
\Lag_\text{int} = y_L \overline\chi \mathcal{S}_L^\dagger F + y_R e^{i\phi_\text{CP}} \overline\chi S_R^\dagger f_R + \hc
\end{align}
where the Yukawa couplings are parametrized by the real constants $y_{L/R}$ and a CP violating phase $\phi_\text{CP}$. In what follows we will focus for simplicity in the case where the dark matter is leptophilic, so that the SM doublet is $F = (\nu_L, f_L)$, with $f_L$ being the left component of a SM lepton and $\nu_L$ the corresponding neutrino. Accordingly, $\mathcal{S}_L=(S^0_L, S^-_L)$ and $S_R=S^-_R$.

The scalar particles of the model ($S_L$, $S_R$ and the SM Higgs boson $\Phi$) interact with one another via trilinear and quartic interactions of the form:
\begin{align} 
    -\Lag_\text{trilinear} =&  A (\mathcal{S}^\dagger_L \Phi) S_R +{\rm h.c.} , \nonumber \\
    -\Lag_\text{quartic} =  &\sum_{S=\mathcal{S}_L, S_R} \frac{1}{2} \lambda_0^S (S^\dagger S)^2 + \lambda_1^S (\Phi^\dagger \Phi) (S^\dagger S) \nonumber \\
    &+\lambda_2^{\mathcal{S}_L} (\Phi^\dagger \mathcal{S}_L)(\mathcal{S}_L^\dagger \Phi) + \lambda_3^{\mathcal{S}_L S_R} (\mathcal{S}_L^\dagger \mathcal{S}_L)(S_R^\dagger S_R),
\end{align}
where $A$ is a constant with mass dimension +1. 

After electroweak symmetry breaking, the charged scalars 
 $(S^-_L,S^-_R)$ mix with one another. The mass eigenstates $(S_1,S_2)$ are constructed through the   transformation:
\begin{align}
    \begin{pmatrix}
    S_L\\
    S_R
    \end{pmatrix}=
    \begin{pmatrix}
    \cos\psi & -\sin\psi\\
    \sin\psi &\cos\psi
    \end{pmatrix}
    \begin{pmatrix}
    S_1\\
    S_2
    \end{pmatrix},
\end{align}
so that the mass term of the Lagrangian is diagonal and of the form:
\begin{equation}
  \Lag_\text{scalar}^\text{mass} = -m_{S_L^0}^2 (S_L^0)^\dagger S_L^0 -m_{S_1}^2 S_1^\dagger S_1  - m_{S_2}^2 S_2^\dagger S_2.
\end{equation}
In turn, the interaction Lagrangian Eq.~(\ref{eq:toymodel_gauge_interaction}) now reads:
\begin{align}\label{eq:portal_interaction}
    \Lag_\text{portal} 
     =  \bar\chi&\left[y_L \cos\psi P_L + y_R\sin\psi e^{i\phi_\text{CP}} P_R\right] S_1^* f + \hc\nn \\
     \phantom{=} + \bar\chi&\left[-y_L \sin\psi P_L + y_R \cos\psi e^{i\phi_\text{CP}}  P_R\right] S_2^* f + \hc
\end{align}

This interaction generates magnetic- and electric dipole moments, a charge radius, and an anapole moment of $\chi$. Concretely, we find that the magnetic dipole- and electric dipole moment of $\chi$ read,
\begin{align}
    \mu_\chi = -\frac{eQ_f}{32\pi^2m_\chi}\biggr\{&(y_L^2 \cos^2\psi + y_R^2 \sin^2\psi) \mathcal{F}_1\left(\frac{m_f}{m_\chi},\frac{m_{S_1}}{m_\chi}\right)\nn\\
       & +2y_L y_R \cos\psi \sin\psi \cos{\phi_\text{CP}} \mathcal{F}_2\left(\frac{m_f}{m_\chi},\frac{m_{S_1}}{m_\chi}\right)\nn\\
       & + (y_L^2 \sin^2\psi + y_R^2 \cos^2\psi) \mathcal{F}_1\left(\frac{m_f}{m_\chi},\frac{m_{S_2}}{m_\chi}\right)\nn\\
       & -2y_L y_R \cos\psi \sin\psi \cos{\phi_\text{CP}} \mathcal{F}_2\left(\frac{m_f}{m_\chi},\frac{m_{S_2}}{m_\chi}\right) \biggr\},
\end{align}
and
\begin{align}
    d_\chi = \frac{eQ_f}{16\pi^2 m_\chi} y_L y_R \cos\psi\sin\psi \sin\phi_\text{CP}\left[\mathcal{F}_2\left(\frac{m_f}{m_\chi},\frac{m_{S_1}}{m_\chi}\right)-\mathcal{F}_2\left(\frac{m_f}{m_\chi},\frac{m_{S_2}}{m_\chi}\right)\right]
\end{align}
respectively. The explicit expressions of the functions $\mathcal{F}_i$ are given in \cref{sec:one-loop-calculation}. Note that the electric dipole moment vanishes if  \textit{i)} $\sin\phi_\text{CP}=0$, but also when \textit{ii)} $m_{S_1} = m_{S_2}$, or when \textit{iii)} $\cos\psi = 0$, $\sin\psi = 0$, since in these cases, the phase can be absorbed in field redefinitions, and there is no CP violation. 

The anapole moment reads
\begin{align}
    \mathcal{A}_\chi = -\frac{e Q_f}{192\pi^2 m_\chi^2} \biggr\{  &\left[ y_L^2 \cos^2\psi - y_R^2 \sin^2\psi  \right]\mathcal{F}_3 \left(\frac{m_f}{m_\chi},\frac{m_{S_1}}{m_\chi}\right)\nn \\
    + &  \left[ y_L^2 \sin^2\psi - y_R^2 \cos^2\psi  \right]\mathcal{F}_3 \left(\frac{m_f}{m_\chi},\frac{m_{S_2}}{m_\chi}\right) \biggr\},
\end{align}
while the charge radius is
\begin{align}
    b_\chi = \frac{-e Q_f }{384 \pi^2 m_\chi^2}\biggr\{&(y_L^2 \cos^2\psi + y_R^2 \sin^2\psi) \mathcal{F}_4\left(\frac{m_f}{m_\chi},\frac{m_{S_1}}{m_\chi}\right)\nn\\
       & +2y_L y_R \cos\psi \sin\psi \cos{\phi_\text{CP}} \mathcal{F}_5\left(\frac{m_f}{m_\chi},\frac{m_{S_1}}{m_\chi}\right)\nn\\
       & + (y_L^2 \sin^2\psi + y_R^2 \cos^2\psi) \mathcal{F}_4\left(\frac{m_f}{m_\chi},\frac{m_{S_2}}{m_\chi}\right)\nn\\
       & -2y_L y_R \cos\psi \sin\psi \cos{\phi_\text{CP}} \mathcal{F}_5\left(\frac{m_f}{m_\chi},\frac{m_{S_2}}{m_\chi}\right) \biggr\}.
\end{align}
In this model, no DM millicharge is generated at the one-loop level. However, if the dark sector was augmented with an additional dark $U(1)_\text{dark}$ symmetry, the resulting kinetic mixing operator could introduce a DM millicharge, potentially impacting the DM direct detection phenomenology. For Majorana DM, all these multipole moments are zero, except for the anapole moment, which is twice as large as for Dirac dark matter.

To simplify our analysis we will assume henceforth that the scalar mass eigenstate $S_2$ is very heavy and can be integrated out. Furthermore, we will denote the DM couplings to $S_1$ in  Eq.~(\ref{eq:portal_interaction}) as $c_L = y_L \cos\psi$ and $c_R=y_R\sin\psi$, and we will parametrize these two couplings as $c_L = c \cos\theta$ and $c_R = c \sin\theta$, where $c^2=y_L^2 \cos^2\psi+y^2_R\sin^2\psi$ and $\tan\theta=y_R/y_L\tan\psi$. Then, the free parameters of the model are 
$m_\chi, m_{S_1}, c, \sin\theta$ and $\sin\phi_\text{CP}$. Notice that $\sin\theta$ parametrizes the amount of P-violation and $\sin\phi_\text{CP}$ the amount of CP-violation.
Given the high sensitivity of the functions ${\cal F}$ to the relative mass difference between the dark matter mass and mediator mass, for the exploration of the parameter space it will prove convenient to use $\eta-1 = (m_{S_1} - \mDm)/\mDm$ instead of $m_{S_1}$ as a free parameter.

\begin{figure}[t!]
    \centering
        \includegraphics[width=0.49\linewidth]{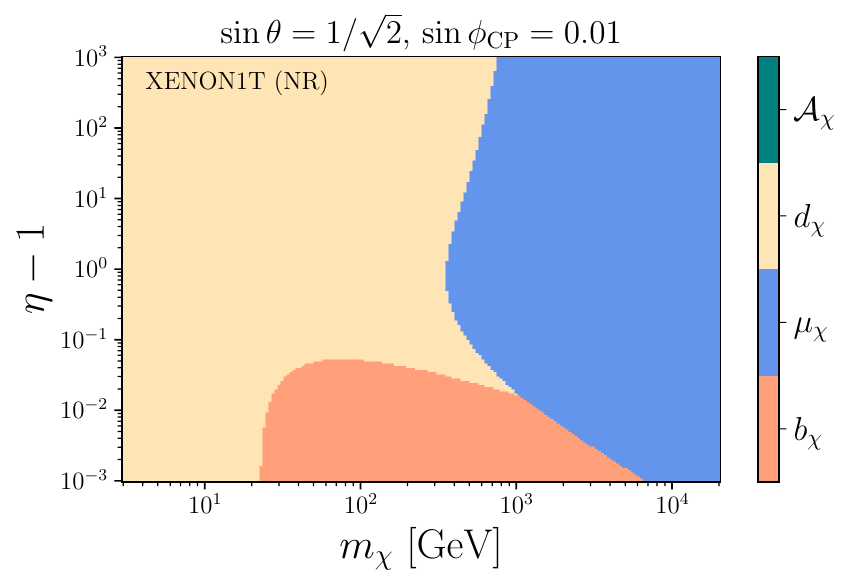}
        \includegraphics[width=0.49\linewidth]{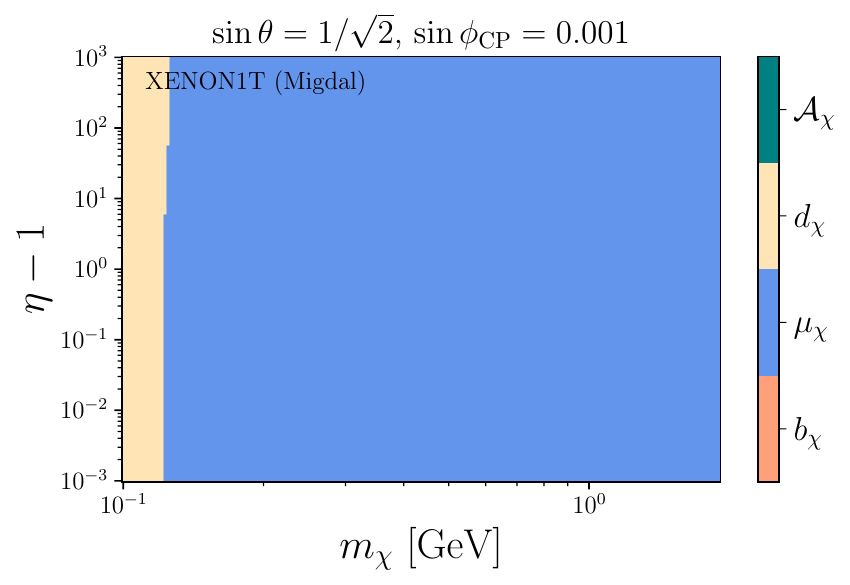}\\
        \includegraphics[width=0.49\linewidth]{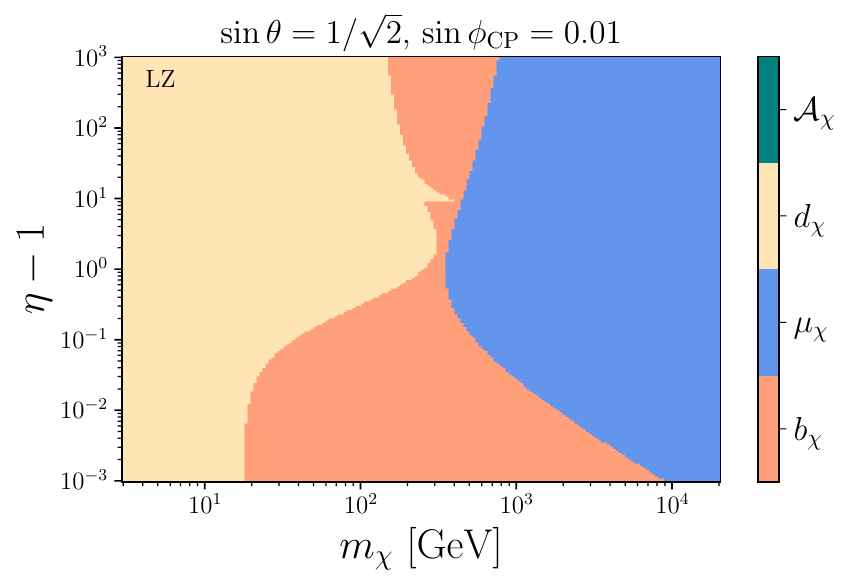}
        \includegraphics[width=0.49\linewidth]{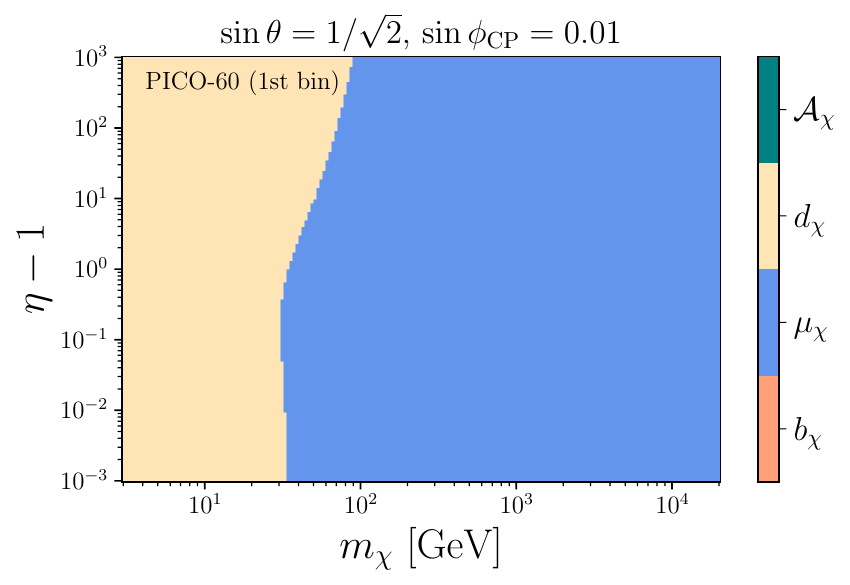}
    \caption{Largest individual contribution to the overall signal rate at the experiments XENON1T (nuclear recoils), XENON1T (Migdal effect), LZ and PICO-60 (1st bin) assuming minimal P-violation ($\sin\theta=1/\sqrt{2}$) and small CP-violation ($\sin\phi_{\rm CP}=0.01$, except for the upper right plot, where we took $\sin\phi_{\rm CP}=0.001$; for $\sin\phi_{\rm CP}=0.01$ the whole parameter space is dominated by $d_\chi$-induced interactions).}
    \label{fig:largest_signal_rate}
\end{figure}

The model-independent constraints derived in section~\ref{subsec:modelin_analysis} can be translated into limits on the parameter space of our simplified model. For concreteness, we consider a tau-philic scenario, \textit{i.e.} $f=\tau$ in Eq.~(\ref{eq:portal_interaction}), and we use Eq.~(\ref{eq:signal_rate_master_eq}) to calculate the rate at an experiment $\mathcal{E}$ for a set of model parameters $\{m_\chi, m_{S_1}, c, \sin\theta, \sin\phi_\text{CP}\}$.
We present in Fig.~\ref{fig:largest_signal_rate} the EM interaction that dominates the total signal rate at a given experiment in the parameter space spanned by $m_\chi$ and $\eta-1$, for representative choices of the  model parameters. We find that all experiments are extremely sensitive to the CP violating angle $\sin\phi_\text{CP}$; even for small values of $\sin\phi_{\rm CP}$,  the electric dipole contribution tends to dominate the overall rate, especially for light dark matter. Furthermore, for those parameters, there is no region of parameter space where the total rate is dominated by the anapole moment (for Dirac dark matter). 

\begin{figure}[t!]
        \includegraphics[width=.49\linewidth]{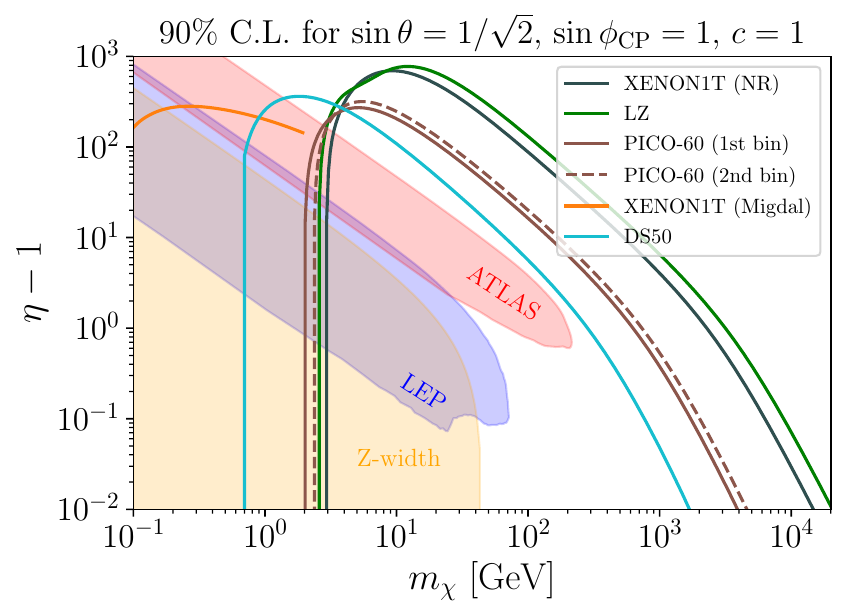}
        \includegraphics[width=0.49\linewidth]{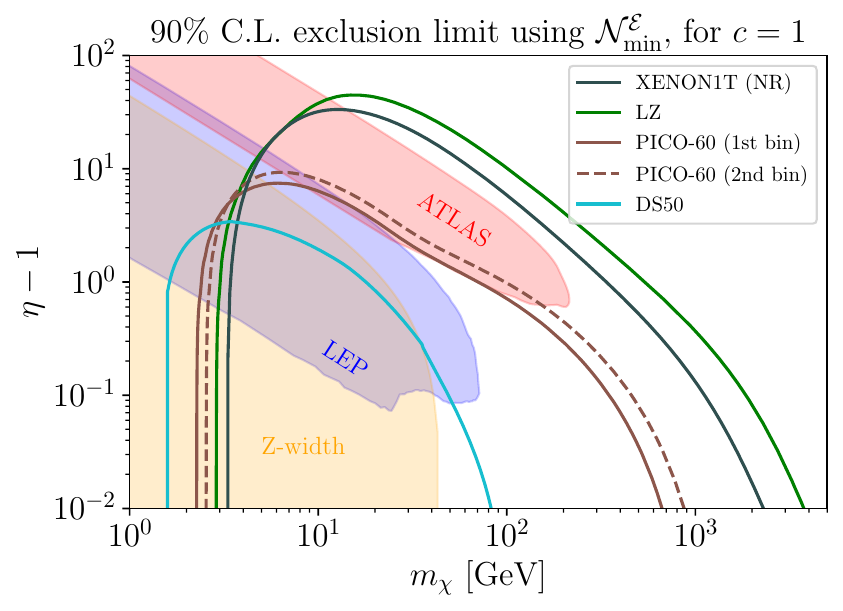}
    \caption{ 90\% C.L. exclusion limits for our tau-philic dark matter scenario, assuming maximal P- and CP violation (left panel) and the most conservative limit sampling over all values of $\sin\theta$ and $\sin\phi_{\rm CP}$ (right panel).
}
    \label{fig:exclusion_limits_Nmin_and_CP}
\end{figure}

In Fig. \ref{fig:exclusion_limits_Nmin_and_CP} we show the 90\% C.L. exclusion limits in the parameters space spanned by the DM mass $\mDm$ and the mass splitting $\eta-1 = (m_{S_1} - \mDm)/\mDm$, for $c=1$ and for two different scenarios. In the left panel, we consider a scenario where the P-violation is minimal and the CP-violation is maximal ({\it i.e.} $\sin\theta=1/\sqrt{2}$, so that $c_L=c_R$, and $\sin\phi_\text{CP}=1$), while in the right panel, we calculate the most conservative limit on the model sampling over $\sin\theta$ and $\sin\phi_\text{CP}$. Namely, we calculate the number of events at an experiment as:
\begin{equation}
\mathcal{N}_\text{min}^{\mathcal{E}}(m_\chi, \eta, c) = \min_{\sin\theta,\sin\phi_\text{CP}} \mathcal{N}_\text{sig}^{\mathcal{E}}(m_\chi, \eta, c, \sin\theta, \sin\phi_\text{CP}).
\label{eq:min_rate}
\end{equation}
Further, the blue and red regions show the constraints on the parameter space from stau searches by LEP \cite{DELPHI:2003uqw} and ATLAS \cite{ATLAS-CONF-2023-029} respectively, while the orange region indicates the region excluded by the $Z$ invisible decay width \cite{ALEPH:2005ab}. For the scenario with minimal P-violation and maximal CP-violation, we find that direct detection experiments are sensitive to rather large mass splittings. Concretely, for $\mDm \simeq 10\gev$, scalar masses $m_{S_1}$ up to  $\sim 10\tev$ could be probed; for larger dark matter masses, $\mDm \simeq \order{1 - 10} \tev$, mass splittings $\eta \simeq \order{1.01 - 1.1}$ could be probed. This large sensitivity of direct detection experiments to our scenario is due to the enhancement of the non-relativistic electric dipole operator in Eq.~(\ref{eq:NR_coefficients}) for small momenta. In the conservative approach, scalar masses up to $\sim 4\tev$ could be probed, surpassing the sensitivity of the ATLAS experiment for $m_\chi\gtrsim 10$ GeV.
We note that xenon-based experiments typically have the best sensitivity for high dark matter masses, whereas in the low mass region, $\mDm \sim 1\gev$, the DS50 experiment provides the best sensitivity.

Additional constraints on our toy model could be imposed from requiring that the thermal freeze-out of our dark matter candidate leads to the observed dark matter abundance $\Omega h^2\simeq 0.12$ \cite{Planck:2018vyg} (see \textit{i.e.} \cite{Garny:2015wea,Ibarra:2015fqa}). On the other, these constraints are highly dependent on the details of the Physics of the Early Universe, and in particular on the possible existence of additional degrees of freedom~\cite{Herms:2021fql}. Similarly, our toy model is also subject to constraints from indirect dark matter searches (see  \textit{i.e.} \cite{Bringmann:2012vr,Garny:2014waa,Garny:2015wea,Sandick:2016zut,Kavanagh:2018xeh}). These constraints are sensitive to uncertainties in the dark matter distribution in the galaxy and will not be further discussed here.

\section{Conclusions}
\label{sec:conclusions}

We have presented a comprehensive analysis of the signatures of a spin 1/2 dark matter particle in a direct detection experiment induced by electromagnetic interactions. We introduced a simple formalism that relates the signal rate at a given direct detection experiment to the dark matter millicharge, charge radius, electric- and magnetic dipole moments, and  anapole moment, including the possible interference among the various interactions.

We have applied this formalism to calculate the number of nuclear recoils induced by dark matter electromagnetic interactions at the XENON1T, LZ, PICO-60 and DS50 experiments. Furthermore, for XENON1T and LZ we also calculated the number of ionizations generated by the Migdal effect. From the non-observation of exotic signals at experiments, we have derived model independent upper limis on the size of the various dark matter electromagnetic multipoles, assuming that only one multipole is present at a time, and also including the possible interference between the millicharge, the charge radius and the magnetic dipole moment. 

Finally, we have considered a simplified dark matter model, consisting of a spin 1/2 dark matter candidate that couples to the left- and right-handed Standard Model leptons through two scalar mediators. The general Lagrangian allows for P and CP-violation, and therefore generates via quantum effects not only a charge radius and a magnetic dipole moment, but also an anapole moment and an electric dipole moment. We have found that the CP violation in this model is strongly constrained by direct detection experiments, due to their high sensitivity to the electric dipole operator. We have also argued the high discovery potential of direct detection experiments, which are able to cover regions of the parameter space which are allowed by collider experiments.

\section*{Note Added}

Simultaneously to our submission, ref~\cite{PandaX:2024pjr} appeared, discussing the search for dark matter electromagnetic multipoles at PandaX-4T experiment.

\acknowledgments

The work of AI and MR was supported by the Collaborative Research Center SFB1258 and by the Deutsche Forschungsgemeinschaft (DFG, German Research Foundation) under Germany's Excellence Strategy - EXC-2094 - 390783311. GT would like to thank Stefano Scopel for useful discussions. MR would like to thank Jaehoon Jeong for helpful discussions.
 
The Feynman diagrams in this paper were drawn with \texttt{TikZ-Feynman} \cite{ELLIS2017103} and evaluated with the help of \texttt{FeynCalc} \cite{MERTIG1991345,Shtabovenko:2016sxi,Shtabovenko:2020gxv}, \texttt{FeynHelpers} \cite{Shtabovenko:2016whf} and \texttt{Package-X} \cite{Patel:2015tea,Patel:2016fam}.

\appendix

\section{Electromagnetic interactions at one-loop}\label{sec:one-loop-calculation}
\begin{figure}[t!]
    \centering
    \begin{subfigure}{.49\linewidth}
        \centering
        \includegraphics{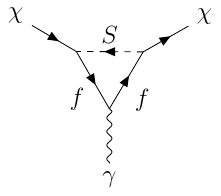}
    \end{subfigure}
    \begin{subfigure}{.49\linewidth}
        \centering
        \includegraphics{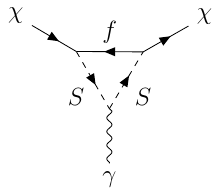}
    \end{subfigure}
    \caption{One-loop diagrams generating dark matter electromagnetic interactions in the simplified t-channel mediator model described by the Lagrangian Eq.~(\ref{eq:LagrangianScalarInteraction}). The DM candidate $\chi$ can be either a Dirac- or a Majorana fermion; for the latter, the conjugated diagrams should also be included.}
    \label{fig:one-loop-diagrams}
\end{figure}

In this appendix, we present the calculation of the electromagnetic interactions generated at the one-loop level in our toy model. We consider the Lagrangian 
\begin{equation}\label{eq:LagrangianScalarInteraction}
\Lag \supset \bar{\chi}\left[c_L P_L + e^{i\phi_\text{CP}}c_R P_R\right] S^*f +\hc,
\end{equation}
where $\chi$ is a Majorana- or Dirac DM candidate, $P_{L/R}$ are the left/right projectors, $S$ is a scalar charged under $U(1)_\text{em}$ and $f$ is a SM fermion. The real Yukawa couplings $c_{L/R}$ characterize the interaction strength to the different chiralities of the SM fermions, and $\phi_\text{CP}$ is a possible CP violating phase. This interaction Lagrangian leads to an interaction with the photon through the diagrams shown in Fig.~\ref{fig:one-loop-diagrams}.

In the following we present the analytical expressions for the electromagnetic moments generated by the t-channel portal Eq.~\ref{eq:LagrangianScalarInteraction}.

\begin{figure}	
        \includegraphics[width=0.49\linewidth]{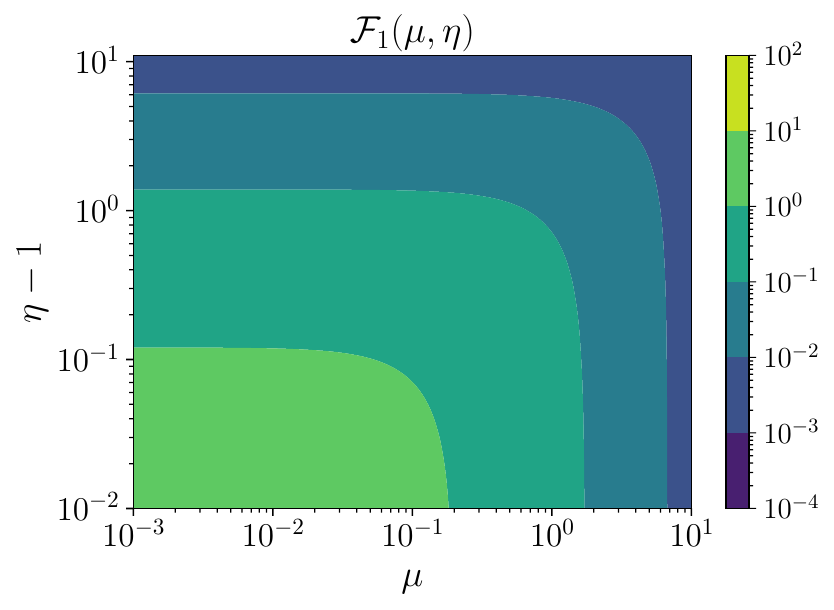}
        \includegraphics[width=0.49\linewidth]{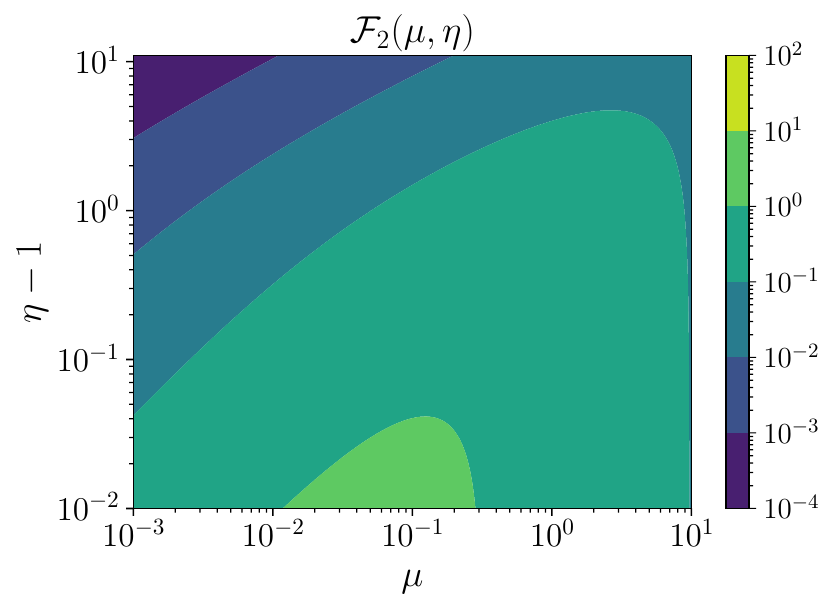} \\
        \includegraphics[width=0.49\linewidth]{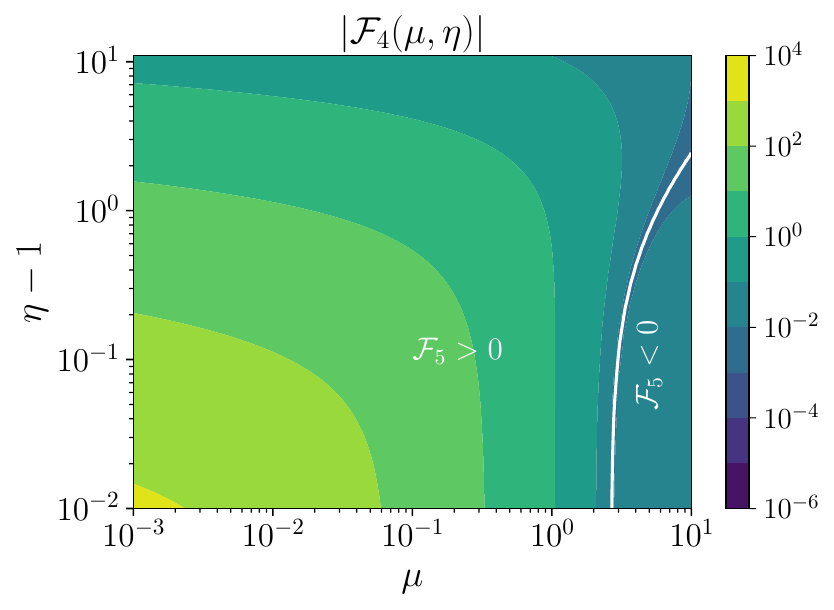}
        \includegraphics[width=0.49\linewidth]{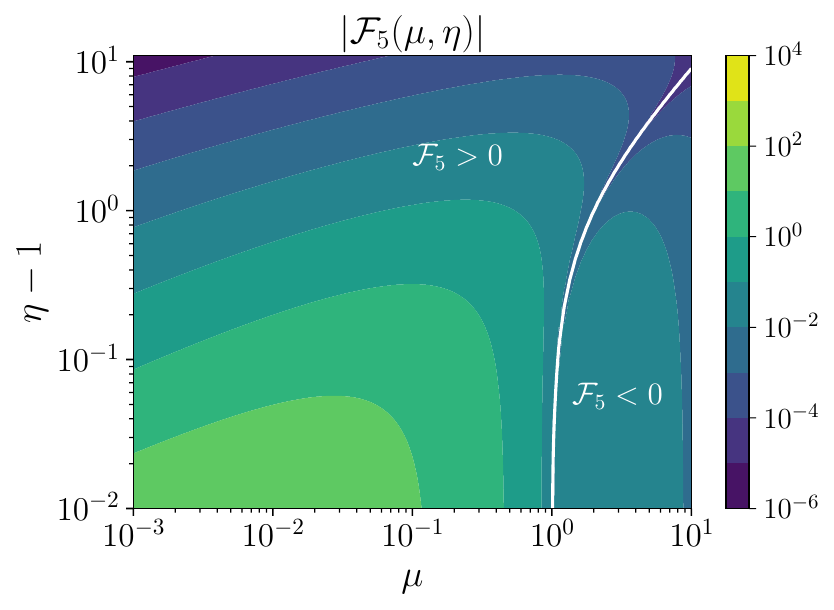}
    \caption{Behavior of the functions $\mathcal{F}_{1,2}(\mu,\eta)$ (top panels) and $|\mathcal{F}_{4,5}(\mu,\eta)|$ (bottom panels) appearing in the dipole moments of Dirac fermions generated by a t-channel scalar mediator. Note that we use different color schemes for the top- and lower figures.}
    \label{fig:F_functions}
\end{figure}

\paragraph{Magnetic Dipole Moment:}
We find that the magnetic moment is given by:
\begin{equation}
\mu_\chi = \frac{-eQ_f}{32\pi^2m_\chi}\biggr\{(c_L^2+c_R^2)\,\mathcal{F}_1 \left(\frac{m_f}{m_\chi},\frac{m_{S}}{m_\chi}\right)+2 c_L c_R \cos\phi_\text{CP}\,\mathcal{F}_2\left(\frac{m_f}{m_\chi},\frac{m_{S}}{m_\chi}\right)\biggr\},
\end{equation}
where $Q_f$ denotes the charge of the internal fermion in units of positron charge $e>0$. The loop functions are defined as
\begin{align}
\mathcal{F}_1(\mu,\eta) =&-1+\frac{1}{2}(\mu^2-\eta^2)\log(\frac{\mu^2}{\eta^2})\nn\\
& -\frac{(\eta^2-1)(\eta^2-2\mu^2)-\mu^2(3-\mu^2)}{\sqrt{\Delta}}\arctanh{\frac{\sqrt{\Delta}}{\eta^2+\mu^2-1}}
\end{align}
and
\begin{align}
\mathcal{F}_2(\mu, \eta) = \mu \left[\frac{1}{2}\log(\frac{\mu^2}{\eta^2})+\frac{\eta^2-\mu^2+1}{\sqrt{\Delta}}\arctanh{\frac{\sqrt{\Delta}}{\eta^2+\mu^2-1}}\right],
\end{align}
with $\Delta = (\mu^2-\eta^2+1)^2-4\mu^2$. We show the behavior of the loop functions in the top panels of Fig.~\ref{fig:F_functions}.
In the limit $m_\chi \ll m_S$ these functions reduce to 
\begin{align}
\mathcal{F}_1(\mu,\eta) &\simeq \frac{\eta^4-\mu^4+2\eta^2\mu^2\log(\mu^2/\eta^2)}{2(\eta^2-\mu^2)^3},\\
\mathcal{F}_2(\mu,\eta) &\simeq \mu\frac{\mu^2-\eta^2-\eta^2\log(\mu^2/\eta^2)}{(\eta^2-\mu^2)^2}.
\end{align}
By expressing the Yukawa couplings as $c_L = c \cos\psi$ and $c_R=c \sin\psi$, we can identify the extrema of the magnetic moment as
\begin{align}
|\mu^\text{max}_\chi| &= \frac{e |Q_f| c^2}{32 \pi^2 m_\chi}\left[ \abs{\mathcal{F}_1\left(\frac{m_f}{m_\chi},\frac{m_{S}}{m_\chi}\right)} + \abs{\mathcal{F}_2\left(\frac{m_f}{m_\chi},\frac{m_{S}}{m_\chi}\right)} \right]\\
|\mu^\text{min}_\chi| &= \frac{e |Q_f| c^2}{32 \pi^2 m_\chi}\times
\begin{cases}
\abs{\abs{\mathcal{F}_1\left(\frac{m_f}{m_\chi},\frac{m_{S}}{m_\chi}\right)} - \abs{\mathcal{F}_2\left(\frac{m_f}{m_\chi},\frac{m_{S}}{m_\chi}\right)} } & \text{if}\,\,|\mathcal{F}_1| > |\mathcal{F}_2|\\
0 &\text{else.}
\end{cases}
\end{align}

\paragraph{Electric Dipole Moment:}

For the electric moment we find
\begin{equation}
d_\chi =\frac{eQ_f}{16\pi^2m_\chi} c_L c_R \sin\phi_\text{CP} \mathcal{F}_2\left(\frac{m_f}{m_\chi}, \frac{m_S}{m_\chi}\right).
\end{equation}

As expected, if the theory preserves CP, e.g. the couplings are real ($\phi_\text{CP}=0$), the electric dipole moment vanishes. Similar to the magnetic dipole case, the functional behavior of $\mathcal{F}_2(\mu,\eta)$ leads to an upper limit on the theory prediction of the electric dipole moment
\begin{equation}
|d_\chi^\text{max}| = \frac{e Q_f c^2}{32 \pi^2 m_\chi} \abs{\mathcal{F}_2\left(\frac{m_f}{m_\chi},\frac{m_{S}}{m_\chi}\right)}.
\end{equation}

In the limit of $m_\chi\ll m_{S}$ the electric dipole moment reduces to,
\begin{equation}
d_\chi\simeq \frac{eQ_f }{16\pi^2} c_L c_R \sin\phi_\text{CP} \frac{\sqrt{\rho}(\rho-1-\log\rho)}{m_S (\rho-1)^2}
\end{equation}
with $\rho=m_f^2/m_S^2$.

\paragraph{Anapole Moment:}

For the anapole moment the result reads:
\begin{equation}\label{Anapole:eq:Fscalar}
\mathcal{A}_\chi =- \frac{e Q_f}{192\pi^2\mDm^2}\left[c_L^2-c_R^2\right]\mathcal{F}_3\Big(\frac{m_f}{m_\chi},\frac{m_S}{m_\chi}\Big)\;
\end{equation}
with
\begin{equation}
\mathcal{F}_3(\mu,\eta)=\frac{3}{2}\log(\frac{\mu^2}{\eta^2})+\frac{3\eta^2-3\mu^2+1}{\sqrt\Delta} \arctanh{\frac{\sqrt{\Delta}}{\eta^2+\mu^2-1}}.
\end{equation}
For $m_\chi \ll m_S$ this reduces to
\begin{equation}
\mathcal{F}_3(\mu,\eta)\simeq \frac{3(\eta^2-\mu^2)+(2\eta^2+\mu^2)\log\frac{\mu^2}{\eta^2}}{(\eta^2-\mu^2)^2}.
\end{equation}
The maximal anapole moment reads
\begin{equation}
|\mathcal{A}_\chi^\text{max}| = \frac{e |Q_f|c^2}{96 \pi^2 m_\chi^2} \abs{\mathcal{F}_3\left(\frac{m_f}{m_\chi},\frac{m_{S}}{m_\chi}\right)}.
\end{equation}

\paragraph{Charge radius}

We find for the charge radius operator
\begin{equation}
b_\chi = \frac{-e Q_f }{384 \pi^2 m_\chi^2}\left[(c_L^2+c_R^2) \mathcal{F}_4\left(\frac{m_f}{m_\chi},\frac{m_{S}}{m_\chi}\right) + 2 c_L c_R \cos\phi_\text{CP} \mathcal{F}_5\left(\frac{m_f}{m_\chi},\frac{m_{S}}{m_\chi}\right) \right],
\end{equation}
where we defined
\begin{align}
\mathcal{F}_4(\mu,\eta) &= \frac{2\left( 8\Delta^2 + \Delta(9\eta^2+7\mu^2-5)-4\mu^2(3\eta^2+\mu^2-1)\right)}{\Delta^{3/2}} \arctanh{\frac{\sqrt{\Delta}}{\eta^2+\mu^2-1}} \nn \\
&+\frac{4(4\Delta+\eta^2+3\mu^2-1)}{\Delta} + (8\mu^2-8\eta^2-1)\log(\frac{\eta^2}{\mu^2})
\end{align}
and
\begin{align}
\mathcal{F}_5(\mu,\eta)  = 8 \mu \biggr[&\frac{\Delta+\eta^2 (-\Delta+2 \mu^2+1)+\mu^2 (\Delta-2 \mu^2+3)-1 }{\Delta^{3/2}} \arctanh{\frac{\sqrt{\Delta}}{\eta^2+\mu^2-1}}\nn \\
&+ \frac{\mu^2-\eta^2}{\Delta} + \frac{1}{2} \log(\frac{\eta^2}{\mu^2}) \biggr].
\end{align}
In the bottom panels of Fig.~\ref{fig:F_functions} we present the functional behaviour of $\mathcal{F}_{4}$ and $\mathcal{F}_5$.
These functions reduce in the limit of $m_\chi \ll m_S$ to 
\begin{align}
\mathcal{F}_4(\mu,\eta) &\simeq -2\frac{3 (\eta^2 - \mu^2) + (2 \eta^2 + \mu^2) \log\left(\mu^2/\eta^2\right)}{(\eta^2 - \mu^2)^2},\\
\mathcal{F}_5(\mu,\eta) &\simeq -4\mu \frac{3(\eta^4-\mu^4)+(\eta^4+4\eta^2\mu^2+\mu^4)\log(\mu^2/\eta^2)}{(\eta^2-\mu^2)^4}.
\end{align}

We find for the extrema,
\begin{align}
|b_\chi^\text{max}| &= \frac{e|Q_f|c^2}{384\pi^2 m_\chi^2} \left[ \abs{\mathcal{F}_4\left(\frac{m_f}{m_\chi},\frac{m_{S}}{m_\chi}\right)} + \abs{\mathcal{F}_5\left(\frac{m_f}{m_\chi},\frac{m_{S}}{m_\chi}\right)}\right] \\
|b_\chi^\text{min}| &= \frac{e|Q_f|c^2}{384\pi^2 m_\chi^2}\times
\begin{cases}
\abs{\abs{\mathcal{F}_4\left(\frac{m_f}{m_\chi},\frac{m_{S}}{m_\chi}\right)} - \abs{\mathcal{F}_5\left(\frac{m_f}{m_\chi},\frac{m_{S}}{m_\chi}\right)} } & \text{if}\,\,|\mathcal{F}_4| > |\mathcal{F}_5|\\
0 &\text{else.}
\end{cases}
\end{align}

\begin{figure}	
\centering
        \includegraphics[width=0.49\linewidth]{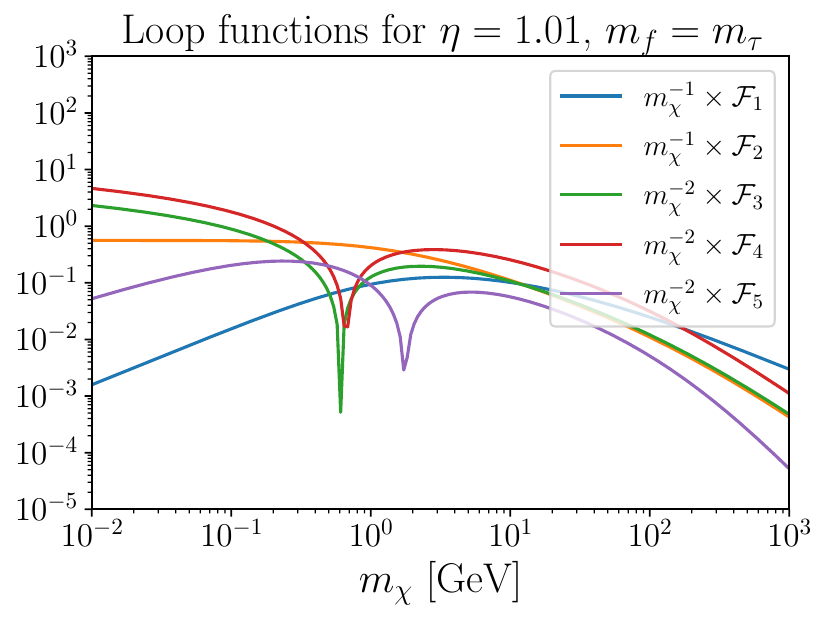}
        \includegraphics[width=0.49\linewidth]{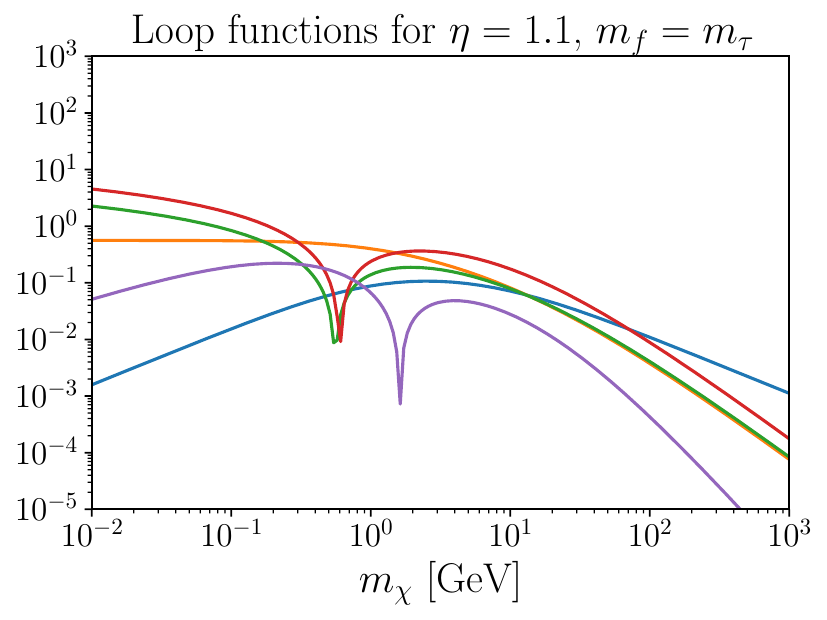}\\
        \includegraphics[width=0.49\linewidth]{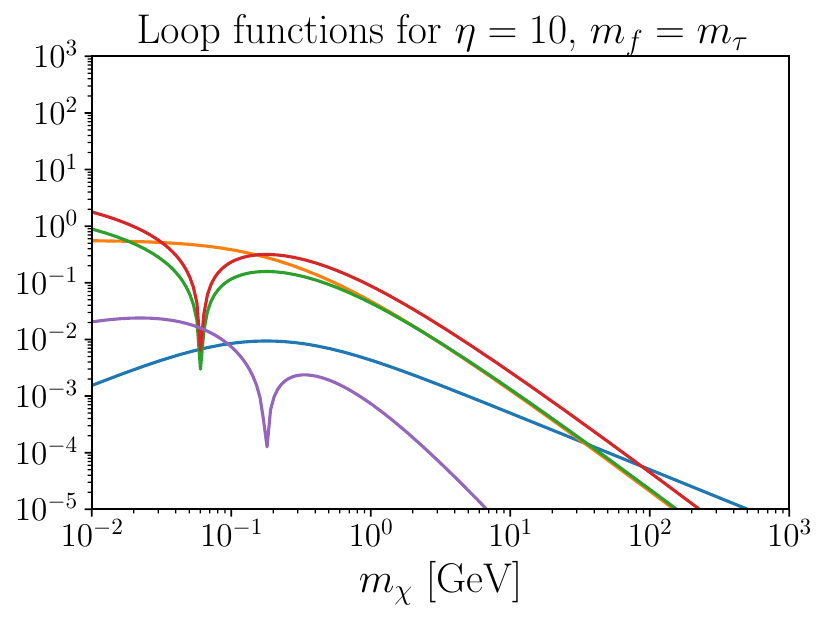}
        \includegraphics[width=0.49\linewidth]{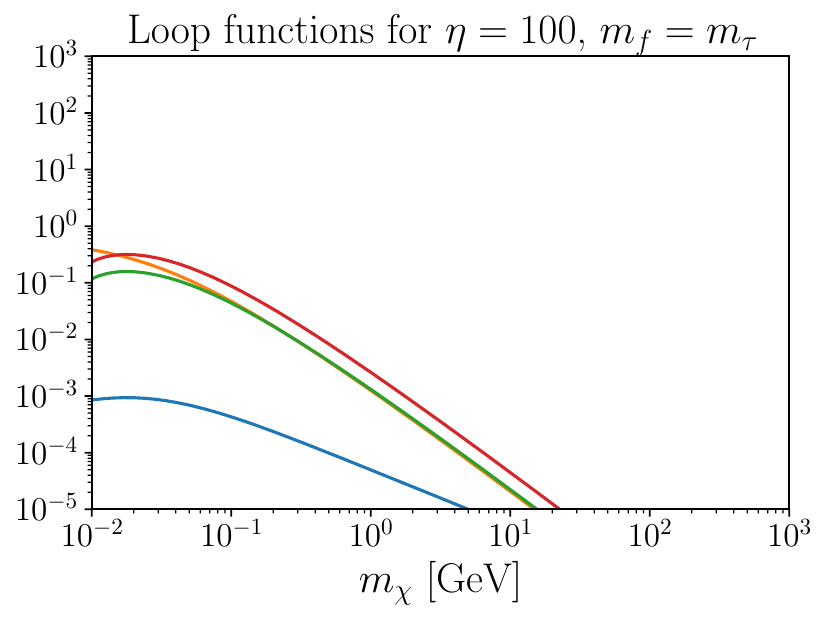}
    \caption{Values of the loop functions $\mathcal{F}_i$ as a function of the dark matter mass, for $m_f = m_\tau$ and for different values of $\eta = m_S/m_\chi$. For convenience in the presentation, the loop functions are multiplied by a power of the dark matter mass, as indicated in the legend of the top-left panel.}
    \label{fig:all_Fs}
\end{figure}

Our results are in agreement with similar studies conducted in Refs.\,\cite{Kopp:2014tsa,Garny:2015wea,Ibarra:2015fqa,Sandick:2016zut,Hisano:2018bpz,Baker:2018uox,Herrero-Garcia:2018koq,Ibarra:2022nzm}. We further present the loop functions in Fig.~\ref{fig:all_Fs} as a function of DM mass $\mDm$ for fixed $\mu = m_\tau/m_\chi$, highlighting their enhancement for small $\eta$.
\newpage
\bibliographystyle{JHEP}

\providecommand{\href}[2]{#2}\begingroup\raggedright\endgroup

\end{document}